\documentclass{elsarticle}
	\makeatletter
\let\c@author\relax
\makeatother

\expandafter\let\csname ver@natbib.sty\endcsname\relax

\usepackage[utf8]{inputenc}
\usepackage{appendix}
\usepackage{graphics,epsfig,amssymb}
\usepackage{amsmath}
\usepackage{amsthm}
\usepackage{amsfonts}
\usepackage{amssymb,latexsym}
\usepackage[all]{xy}
\usepackage{graphicx,color}
\usepackage[mathscr]{eucal}
\usepackage{float}

\usepackage{enumitem} 
\usepackage{xcolor}
\usepackage{mathtools}
\usepackage{todonotes}
\usepackage{verbatim}
\usepackage{url}
\usepackage{multicol}
\usepackage{colortbl}
\usepackage{booktabs}
\arrayrulecolor{gray}
\usepackage{subcaption}

\usepackage[commandnameprefix=ifneeded]{changes}
\usepackage{verbatim}

\usepackage{tikz}

\usepackage{hyperref}
\hypersetup{
    colorlinks=true,
    linkcolor=blue,
    filecolor=magenta,      
    urlcolor=cyan,
    pdftitle={Overleaf Example},
    pdfpagemode=FullScreen,
    }

\usepackage[backend=biber,style=authoryear, isbn=false, url=false,date=year,giveninits=true,mincitenames=2,minbibnames=2]{biblatex}
\DeclareDelimFormat{nameyeardelim}{\addcomma\space}
\usetikzlibrary{decorations.pathreplacing}
\usetikzlibrary{patterns}
\usetikzlibrary{positioning,arrows,arrows.meta}
\renewbibmacro{in:}{%
  \ifentrytype{article}{}{\printtext{\bibstring{in}\intitlepunct}}}
\DeclareFieldFormat[article,inbook,incollection,inproceedings,patent,thesis,unpublished]  {title}{#1\isdot}
\AtEveryBibitem{% Clean up the bibtex rather than editing it 
 \ifentrytype{online}{}{% Remove publisher and editor except for books
  \clearfield{note}
 }
}
\usepackage{pgfplots}
\usepgfplotslibrary{fillbetween}
\newtheorem{theorem}{Theorem}[section]

\newtheorem{lemma}[theorem]{Lemma}
\newtheorem{proposition}[theorem]{Proposition}

\theoremstyle{definition}

\theoremstyle{definition}

\theoremstyle{definition}

\def\E{{\mathbb E}}

\def\V{{\mathbb V}}
\newcommand{\Ro}{\mathcal{R}_0}
\newcommand{\dd}{\textnormal{d}}
\newcommand{\pcite}[1]{\parencite{#1}}
\newcommand{\tcite}[1]{\textcite{#1}}

\makeatletter
\g@addto@macro{\UrlBreaks}{\UrlOrds}
\makeatother

\graphicspath{{./}{./figures/}}

\begin{document}

\begin{frontmatter}

\title{SEIR models with host heterogeneity:\\ theoretical aspects and applications to seasonal influenza dynamics}
\author[add1,add2]{Tam\'as Tekeli}
\ead{tekeli@math.u-szeged.hu}
\author[add1]{Andrea Pugliese\corref{cor1}
}
\ead{andrea.pugliese@unitn.it}
\author[add1]{Cinzia Soresina}
\ead{cinzia.soresina@unitn.it}
\cortext[cor1]{Corresponding author}
\affiliation[add1]{organization={Department of Mathematics, University of Trento},
addressline={via Sommarive 14},
postcode={38123},
city={Povo, Trento},
country={Italy}}
\affiliation[add2]{organization={National Laboratory for Health Security, University of Szeged, Bolyai Institute},
addressline={Aradi v\'ertan\'utere 1},
postcode={6720},
city={Szeged},
country={Hungary}}
\begin{abstract}
  Population heterogeneity is a key factor in epidemic dynamics, influencing both transmission and final epidemic size. While heterogeneity is often modelled through age structure, spatial location, or contact patterns, differences in host susceptibility have recently gained attention, particularly during the COVID-19 pandemic. 
Building on the framework of Diekmann and Inaba (Journal of Mathematical Biology, 2023), we focus on the special case of SEIR epidemic models, assuming that at the epidemic start there is no pre-existing immunity. Under two distinct assumptions linking susceptibility and infectiousness, one obtains a closed system of 3 ODEs, which can be easily simulated and for which some analytical results are obtained. 
In particular, we proved that heterogeneity in susceptibility reduces the epidemic final size compared to homogeneous models with the same basic reproduction number $R_0$. We specialised in the case where susceptibility is distributed according to a gamma or extended Beta distribution, showing how the epidemic final size depends on the variance of the distribution. In the case of a gamma-distributed susceptibility, the resulting model consists of a system of ODEs with just one parameter more than the classical SEIR model; this makes it practical for fitting epidemic data. We illustrate its use by fitting data on seasonal influenza in Italy, and comparing the results to those obtained with simple SEIR models with pre-existing immunity.

\end{abstract}
\end{frontmatter}

\thispagestyle{plain}

%  Ideas:
% \begin{itemize}

%     \item different distributions
%     \item age-structured analysis
%     \item 
% \end{itemize}
\section{Introduction}
   Population heterogeneity has been recognised as a fundamental determinant of epidemic dynamics. Heterogeneities may be due to age \pcite{AndMay, Rost2020} or spatial location \pcite{Longini_old}, and in this context, the structure of contact matrices \pcite{Polymod} or connections between nodes \pcite{Colizza2007} play an important role;  heterogeneity in the number of contacts has been considered especially in the context of sexually-transmitted infections \pcite{Hym88}, in which case the type of mixing (assortative or proportionate) significantly affects the epidemic dynamics. 
   A different kind of heterogeneity, differences in susceptibility, i.e.\ in the probability of getting infected when contacted by an infectious individual, has been proposed by several authors \pcite{BrittonBallTrapman2020} during the Covid epidemic and systematically developed by \tcite{Novozhilov2008} and, more recently, by \tcite{DiekmannInaba2023} (see also \cite{Bootsma2023} and \cite{Inaba2025}).
   A somewhat similar model had been studied by \tcite{Hyman2005}, who analysed the long-term dynamics of an endemic SIR model in which the population is divided into a discrete number of groups differing in susceptibility to infection.  
   Different approaches to incorporate heterogeneity have been proposed by \tcite{Rose2021} and by \tcite{Berestycki2023}.

   \tcite{DiekmannInaba2023} developed the theory for a general type of epidemic in which infectiousness depends, in an arbitrary way, on the time since infection, and the model is set as a renewal equation; later, they specialise the resulting equation for the case in which the model can be described by a system of ODEs. Here, we consider only the case of SEIR models, which are widely used to describe the dynamics of influenza epidemics, as well as several other infectious diseases. 
   
   %We begin by deriving the equations for this special case, as presented in \tcite{DiekmannInaba2023}, under two different assumptions regarding the relationship between susceptibility and infectiousness. We believe that focusing on this specific case, the derivation is easier to follow, although the systems obtained have already been presented in \pcite{DiekmannInaba2023}. 
   The paper is organised as follows. In Section \ref{sec:model}, we derive the equations for this special case, following \tcite{DiekmannInaba2023}, under two alternative assumptions on the relationship between susceptibility and infectiousness. By concentrating on this specific setting, the derivation becomes more transparent, even though the resulting systems have already been presented in \pcite{DiekmannInaba2023}.
   In Section \ref{sec:properties} we study the epidemic final size for the resulting models; there we show, similarly to \tcite{Novozhilov2008}, that the presence of heterogeneity always decreases the epidemic final size relative to a homogeneous model with the same value of $\Ro$, under either assumption on the relation between susceptibility and infectiousness. 
   In Section \ref{sec:distributions}, we specialise the equations to the case where susceptibility heterogeneity follows either a gamma~\pcite{DiekmannInaba2023} or an extended beta distribution. In Section \ref{Sec:finalsize}, we prove that, in the case of gamma distributions, a greater variance in susceptibility lowers the attack ratio, and that, for the same value of $\Ro$, the attack ratio is lower when infectiousness is proportional to susceptibility than when it is independent; we also compare numerically the final size obtained with the gamma distribution with that resulting from beta distributions with the same variance. Finally, in Section~\ref{sec:datainfluenza}, we compare models with and without susceptibility heterogeneity by fitting them to Italian seasonal influenza data from 2014-15 to 2023-24.
   %Then, we consider the case \pcite{DiekmannInaba2023} in which susceptibility heterogeneity follows a gamma distribution; in this case, we show that the higher the variance in susceptibility, the lower the resulting attack ratio. Finally, we compare the behaviour of the models with and without heterogeneity in susceptibility, by fitting both models to data on seasonal influenza in Italy between 2014-15 and 2023-24.  

\section{Model construction}\label{sec:model}
The model assumes homogeneous mixing of the population (although the same ideas can be applied to multi-group or multi-age models with a given contact structure), so that each individual is subject at time~$t$ to a ``force of infection''~$F(t)$, given by the contact rate with infectious individuals. Individuals are characterised by a feature~$x$ from which both susceptibility~$a(x)$ (i.e., the factor multiplying~$F(t)$ to yield the rate at which a susceptible $x$ becomes infected) and infectiousness~$c(x)$ (i.e., the contribution of an infected individual to the force of infection) depend.\\
The feature $x$ could depend on genetic factors, immunological history, or possibly behaviour. Still, it is assumed that it is fixed, at least as long as an epidemic outbreak is concerned, and that its distribution in the population has density $\phi(x)$ (a more general distribution, allowing also for a discrete part, is considered in~\tcite{DiekmannInaba2023}, but here we stick to the simplest case).

Following~\tcite{DiekmannInaba2023}, we can set $a(x)=x$ (one needs only to redefine the feature) and assume that its mean is 1, i.e., $\int_0^\infty x \phi(x)\, \dd x =1$, by redefining $F(t)$ appropriately. 

Considering, as stated in the Introduction, an SEIR ODE model, let now be $s(t,x)$, $e(t,x)$, and $i(t,x)$ the fractions of individuals of type $x$ in the compartments $S$, $E$, and $I$; let the exit rates from the $E$ and $I$ compartments be $\alpha$ and $\gamma$, respectively. From these assumptions, one obtains
\begin{align}
    \frac{\partial}{\partial t} s(t,x) &= - x s(t,x) F(t),\\
     \label{eq_e}\frac{\partial}{\partial t} e(t,x) &=  x s(t,x) F(t) - \alpha e(t,x),\\ \label{eq_i}
    \frac{\partial}{\partial t} i(t,x) &=   \alpha e(t,x) - \gamma i(t,x),\\
    F(t) &= \int_0^\infty c(x) i(t,x) \phi(x) \, \dd x.
\end{align}
The term $\phi(x)$ in the force of infection comes from the assumption that the probability of contacting an individual of type $x$ is proportional to their density in the population.

First of all, from the first equation, one can see that 
\[ s(t,x) = s_0(x) \exp\{-x \int_0^t F(\tau)\, \dd \tau \}.\]
Assuming, in the case of an outbreak, that $s_0(x) \approx 1$ for all $x$, one obtains
\begin{equation}
    \label{s(t,x)}
    s(t,x) = \bar s(t)^x,
\end{equation}
where $\bar s(t) = s(t,1)$. 

The subsequent elaborations depend on the choice of the function $c(x)$; following \tcite{DiekmannInaba2023}, we consider the cases $c(x) \equiv \beta$ (infectiousness independent of susceptibility) and $c(x) = \beta x $ (perfect correlation between susceptibility and infectiousness).

For future use, introduce the functions 
\begin{equation}
    \label{Phi}
    \Phi_k(w) = \int_0^\infty x^k w^x \phi(x)\, \dd x,
\end{equation}
where $k$ may be 0, 1, or 2.

If $c(x) \equiv \beta$ (case named $k$ = 1), one considers the variables
\[ E(t) = \int_0^\infty e(t,x) \phi(x) \, \dd x\]
and
\[ I(t) = \int_0^\infty i(t,x) \phi(x) \, \dd x\]
to obtain $F(t) = \beta I(t)$. Then, integrating \eqref{eq_e}--\eqref{eq_i}, one has the system of three ODEs
\begin{align}
    \bar s'(t) &= - \beta \bar s(t) I(t), \nonumber \\
    E'(t) & = \beta \Phi_1(\bar s(t)) I(t) - \alpha E(t),    \label{sys_bar}
\\
    I'(t) &= \alpha E(t) - \gamma I(t). \nonumber
 \end{align}

If $c(x) \equiv \beta x$ (case named $k$ = 2), one defines
\[ U(t) = \int_0^\infty x \,e(t,x) \phi(x) \, \dd x\]
and
\[ V(t) = \int_0^\infty x\, i(t,x) \phi(x) \, \dd x\]
to obtain $F(t) = \beta V(t)$.

Now, integrating \eqref{eq_e}--\eqref{eq_i}, one obtains
\begin{align}
    \bar s'(t) &= - \beta \bar s(t) V(t), \nonumber \\
    U'(t) & = \beta \Phi_2(\bar s(t)) V(t) - \alpha U(t),    \label{sys_bar2}
\\
    V'(t) &= \alpha U(t) - \gamma V(t). \nonumber
 \end{align}
While \eqref{sys_bar2} is a closed system, it may be convenient to also consider the variables $E(t)$ and $V(t)$, in order to compare the model results with data.
One has
\begin{align}
    E'(t) & = \beta \Phi_1(\bar s(t)) V(t) - \alpha E(t), \nonumber \\[-0.5em] \   \label{sys_bar2b}
\\[-0.5em]
    I'(t) &= \alpha E(t) - \gamma I(t). \nonumber
 \end{align}

Although the function $\Phi_k$ cannot generally be explicitly computed (a special case is the gamma distribution for $x$, see Section~\ref{subsec:gamma-dis}), there is no problem in numerically integrating system \eqref{sys_bar}.

\section{Basic properties}\label{sec:properties}
One can easily see some properties of systems \eqref{sys_bar} and \eqref{sys_bar2}. First of all, linearising system \eqref{sys_bar} for $E(t)$ and $I(t)$, or system \eqref{sys_bar2} for $U(t)$ and $V(t)$, in both cases assuming $\bar s(t) \equiv 1$, one sees that an epidemic outbreak can start if and only if $\beta \Phi_k(1) > \gamma$, where $k=1$ or $2$ corresponds to the model used. Equivalently, the condition for an epidemic outbreak can be written as  $\mathcal{R}_0 > 1$, defining naturally
\begin{equation}
\label{R0}
    \mathcal{R}_0 = \frac{\beta \Phi_k(1)}{\gamma}.
\end{equation}
In case $k=1$, $\Phi_1(1)=1$ and the definition of $\mathcal{R}_0$ is the classical one.

In the case of correlated infectiousness and susceptibilities, $\Phi_2(1)$ involves the variance of the susceptibility $X$, since $\Phi_2(1) = 1 + {Var(x)}$. The resulting formula is similar to that obtained with the proportionate mixing in the case of varying contact rates \pcite{AndMay1986}.

% \begin{table}[h!]
% \centering
% \begin{tabular}{ c|c|c} 

% $\mathcal{R}_0$  & $k$ = 1 & $k$ = 2\\
%  \hline
% $\phi(x) \sim Gamma(p)$ & $\tfrac{\beta}{\gamma}$ & $\tfrac{\beta}{\gamma} \cdot \left(1 + \frac{1}{p}\right) \color{red} =   \tfrac{\beta}{\gamma} \cdot \left(1 + Var\right)$\\ \hline
% $\phi(x) \sim Beta(a, b, 0, L)$ & $\tfrac{\beta}{\gamma}$ & $\tfrac{\beta}{\gamma} \cdot \frac{(a + b)(a+1)}{a(a+b+1)} \color{red} = \tfrac{\beta}{\gamma} \cdot \left(1 + Var\right)$ \\ 
%  \hline

% \end{tabular}
% \caption{Formulae for $\mathcal{R}_0$ considering differently distributed heterogeneity.}
% \end{table}

One can also see that the systems admit a prime integral. Indeed, one can easily check that $\Phi_{k-1}(w)$ is a primitive of $\Phi_k(w)/w$.
It then follows that in system \eqref{sys_bar},
$$\dfrac{\dd}{\dd t}\left(\Phi_{0}(\bar s(t)) + E(t) + I(t) - \dfrac{\gamma}{\beta}\log(\bar s(t))\right) = 0.$$
Similarly, in system \eqref{sys_bar2},
$$\dfrac{\dd}{\dd t}\left(\Phi_{1}(\bar s(t)) + U(t) + V(t) - \frac{\gamma}{\beta}\log(\bar s(t))\right) = 0.$$

Moreover, it is not difficult to show using standard methods that necessarily there exists $\bar s_\infty = \lim\limits_{t \to +\infty} \bar s(t)$ and that
\[ \lim_{t \to +\infty} E(t) = \lim_{t \to +\infty} I(t)= \lim_{t \to +\infty} U(t)= \lim_{t \to +\infty} V(t) = 0.\]
Then, assuming that an epidemic has started with $\bar s_0 \approx 1$ and $E_0 \approx I_0 \approx 0$, from the prime integrals, one obtains
\begin{equation}
\label{finalsize_gen}
    \Phi_{k-1}(\bar s_\infty) - \frac{\gamma}{\beta}\log(\bar s_\infty)=
    \Phi_{k-1}(\bar s_0) - \frac{\gamma}{\beta}\log(\bar s_0) = 1,
\end{equation}
with $k=1$ for system \eqref{sys_bar}, and $k=2$ for \eqref{sys_bar2}.

\begin{theorem}
    Equation \eqref{finalsize_gen} admits a unique solution $\bar s_\infty \in (0,1)$ if $\Ro > 1$, no solutions in $(0,1)$ if $\Ro \le 1$.\\
   % If $k=1$, 
   The total susceptible fraction $\bar S_\infty = \Phi_0(\bar s_\infty)$ is larger than the final susceptible fraction $S_\infty$ found in the homogeneous model with the same value of $\Ro$.
   \label{theo3.1}
\end{theorem}

The proof can be found in Appendix~\ref{sec:proofs_A}.
In biological terms, Theorem~\ref{theo3.1} shows that with any distribution of heterogeneity in susceptibility, and with both models, the susceptible fraction at the end of an epidemic is larger than with homogeneous susceptibility; in other words, the attack ratio (or epidemic final size) will be smaller. This is similar to the results obtained by~\tcite{Novozhilov2008}.

\section{The system with specific distributions of susceptibility}\label{sec:distributions}
\subsection{The case of gamma-distributed susceptibility}\label{subsec:gamma-dis}
Again following \tcite{DiekmannInaba2023}, we assume that $X$ follows the gamma distribution with mean 1 and variance $1/p$, i.e.,
\begin{equation}
    \label{gamma}
    \phi(x) = \frac{p^p}{\Gamma(p)} x^{p-1}e^{-px}.
\end{equation}
In this case, it is useful to introduce the overall fraction of susceptibles as a variable 
\begin{equation}
    \label{S(t)}
    S(t) = \Phi_0(\bar s(t))=\int_0^{+\infty} \bar s(t)^x \phi(x)\, \dd x = \left(  \frac{p}{p-\log(\bar s(t))}\right)^p.
\end{equation}
Differentiating this expression, we obtain
\[ S'(t) = S(t)^{1+\frac1p} \frac{\bar s'(t)}{\bar s(t)} = - \beta S(t)^{1+\frac1p}  I(t).\] 
To complete the system, we need to express $\Phi_k(\bar s(t))$ in terms of $S(t)$.
A simple computation shows that $\Phi_1(\bar s(t)) = S(t)^{1+\frac1p}$, while $\Phi_2(\bar s(t)) = \left(1+\frac1p \right)S(t)^{1+\frac2p}$. Hence, we have the following cases.
\begin{itemize}
    \item The model in which susceptibility and the infectiousness are not correlated, i.e., $c(x)\equiv \beta$, is 
\begin{align}
S'(t) &= - \beta I(t) \cdot S(t)^{1+\tfrac{1}{p}}, \nonumber\\
E'(t) &= \beta I(t) \cdot S(t)^{1+\tfrac{1}{p}} - \alpha E(t), \label{sys1}\\[0.2cm]
I'(t) &= \alpha E(t) - \gamma I(t). \nonumber 
\end{align}
% R0=beta/gamma, independent of p
% but final size depends on p
    \item The model in which infectiousness is proportional to susceptibility, i.e., $c(x)\equiv \beta x$, is
\begin{align}
S'(t) &= - \beta V(t) \cdot S(t)^{1+\tfrac{1}{p}}, \nonumber \\
U'(t) &= \beta V(t) \cdot \left(1+\tfrac{1}{p}\right)S(t)^{1+\tfrac{2}{p}} - \alpha U(t), \label{sys2}\\[0.2cm]
V'(t) &= \alpha U(t) - \gamma V(t). \nonumber
\end{align}
In this case, we may also wish to consider the equations for $E(t)$ and $I(t)$, which are the same as above
\begin{align}
E'(t) &= \beta V(t) \cdot S(t)^{1+\tfrac{1}{p}} - \alpha E(t),\nonumber\\[-0.5em]
 \label{sys2b}\\[-0.5em]
I'(t) &= \alpha E(t) - \gamma I(t). \nonumber 
\end{align}

\end{itemize}
\begin{comment}

\begin{figure}[H]
	\centering
	\includegraphics[width=\textwidth]{./figures/p0.005.pdf}
    \caption{$p = 0.005, R_0 = \beta \cdot \tfrac{1}{\gamma} = \tfrac{1}{7} = $}
    \includegraphics[width=\textwidth]{./figures/p0.6.pdf}
    \caption{$p = 0.6, R_0 =$}
    \includegraphics[width=\textwidth]{./figures/p5.pdf}
    \caption{$p = 5, R_0 = $}

\end{figure}
 
\textcolor{blue}{Can the peak of the green curve be this overwhelmingly high? The flow from $S$ to $E$ is a bit confusing.}
\end{comment}

\subsection{Beta-distributed susceptibility}
Here we assume that $X$ follows the Beta distribution with shape parameters $a$ and $b$, scaled by a factor $L > 1$, so that its support becomes $(0, L)$ (we will sometimes refer to this as an extended Beta distribution). Then
\begin{equation}
    \label{beta_dist}
    \phi(x) = \begin{cases}
        \dfrac{\left(\dfrac{x}{L} \right)^{a-1}\left(1-\dfrac{x}{L}\right)^{b-1} }{L\,B(a,b)},& x \in (0,L), \\ 0, & \mbox{otherwise},
    \end{cases}
\end{equation}
where $B(a,b) = \int_0^1 x^{a-1}(1-x)^{b-1}\, \dd x$.

It is well known that, for a standard Beta random variable $X$, 
$$\E(X) = \dfrac{a}{a+b},\quad \V(X) = \dfrac{a b }{(a+b)^2(a+b+1)}.$$ Multiplying $X$ by $L$, one then sees that the mean susceptibility is equal to 1 and its variance is equal to $V$ if the following relations hold
\begin{align}
    \frac{a L }{a+b} = 1, \qquad
    \frac{b}{a \left(a+b+1\right)} = V.
    \label{L_beta}
\end{align}
It is clear that, for a given variance $V$, there are infinitely many combinations of $a$, $b$, and $L$ satisfying \eqref{L_beta}.

For instance, one can choose any value of $a \in (0,1/V)$ and then set
\begin{align}
   L = \frac{1+V}{1-a V}, \qquad
    b = \frac{a(1+a) V}{1 - a V}.
    \label{choice_beta1}
\end{align}
Notice that, for a given value of $a$, if one lets $V$  increase to $1/a$ (the largest possible value) while using \eqref{choice_beta1}, the distribution converges to a Gamma$(a)$.

Alternatively, one can choose any value of $L > V+1 $ and then set
\begin{equation}
    a = \frac{L-V-1}{LV},\qquad b = \frac{(L-V-1)(L-1)}{LV}.\label{choice_beta2}
\end{equation} 
Notice that, when $a=b=1$, and then necessarily $L=2$, one obtains the uniform distribution, which is thus a special case of the Beta.\\
With this parametrization, if $L$ is fixed, and $V$  increases to $L-1$ (the largest possible value) while using \eqref{choice_beta2}, the distribution converges to a bimodal distribution where $X=L$ with probability $1/L$ and $X=0$ with probability $(L-1)/L$.

When $\phi$ is given by \eqref{beta_dist}, the function $\Phi_k$ defined in \eqref{Phi} can be obtained as
\begin{equation}
    \label{Phi_beta}
    \Phi_k(w)= \frac{L^k}{B(a,b)} \int_0^1 e^{L \log(w) x} x^{k+a-1}(1-x)^{b-1} \, \dd x.
\end{equation}
This expression can be rewritten in terms of Kummer's function $M(a,b,z)$ \autocite[p.\ 505]{Abramowitz}, also known as the hypergeometric confluent function ${}_1F_1(a,b,z)$  whose numerical computation is provided in several software packages (e.g.\ Wolfram-Alpha, Python, or Matlab), making computations much faster. Precisely, one has
\begin{equation}
    \label{Phi_beta2}
    \Phi_k(w)= \frac{L^k\Gamma(a+b)\Gamma(a+k)}{\Gamma(a)\Gamma(a+b+k)} M(a+k,a+b+k,L\log(w)).
\end{equation}

\subsection{Numerical illustrations}
\begin{figure}
\centering
\begin{tabular}{cc}
   \includegraphics[width=0.45\textwidth]{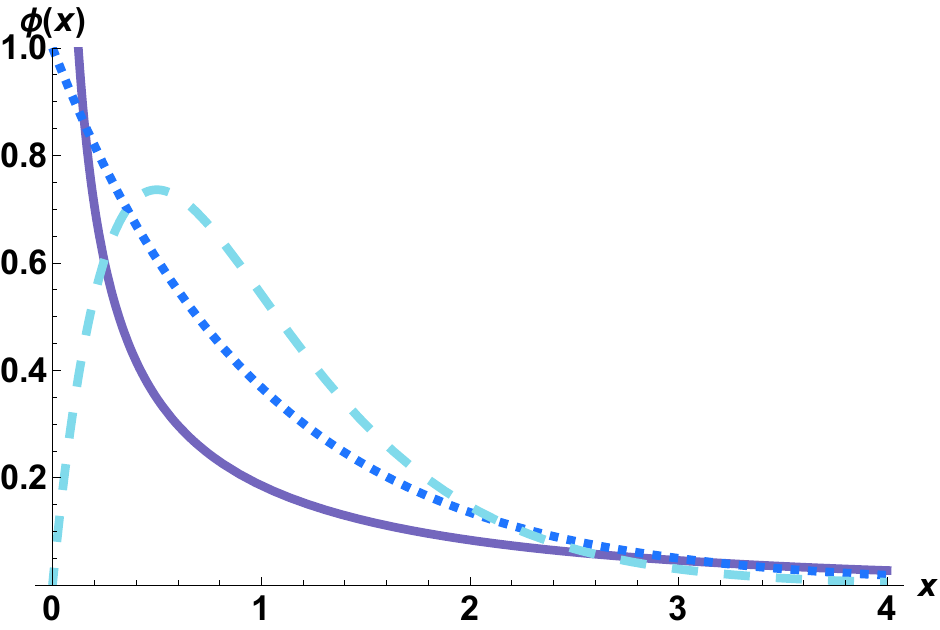}  &   \includegraphics[width=0.45\textwidth]{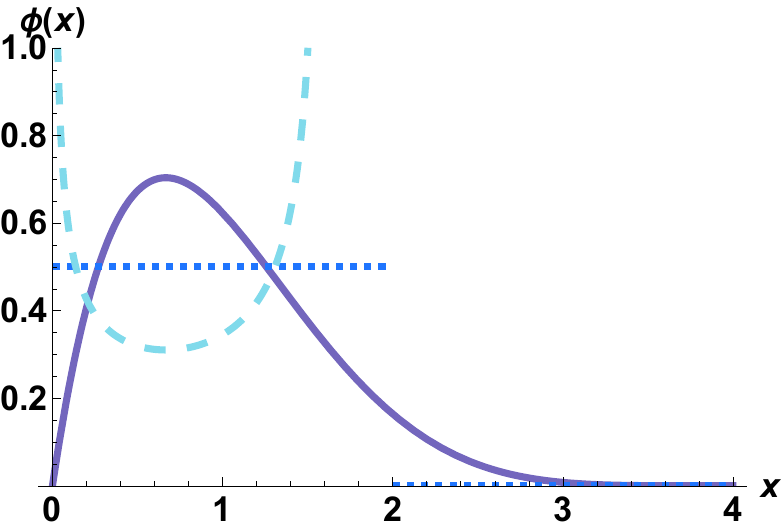}
    \\
 \includegraphics[width=0.47\textwidth]{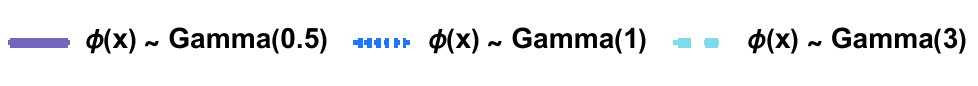}
      &   \includegraphics[width=0.47\textwidth]{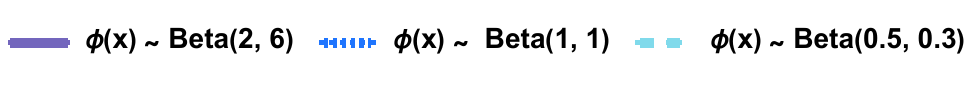}
   \end{tabular}
    \caption{Plots of Gamma (left) and Beta (right) susceptibility distributions used in the simulations in Figure \ref{fig1}. gamma distributions have variance $1/p$, where $p$ is specified in the legend. For the Beta distributions, the values of $a$ and $b$ are also specified in the legend, while $L$ is obtained from \eqref{L_beta}; it can be verified that all Beta distributions have variance 1/3.  }
\label{fig0}
\end{figure}

\begin{figure}%[H]
	\centering
	\includegraphics[width=0.45\textwidth]{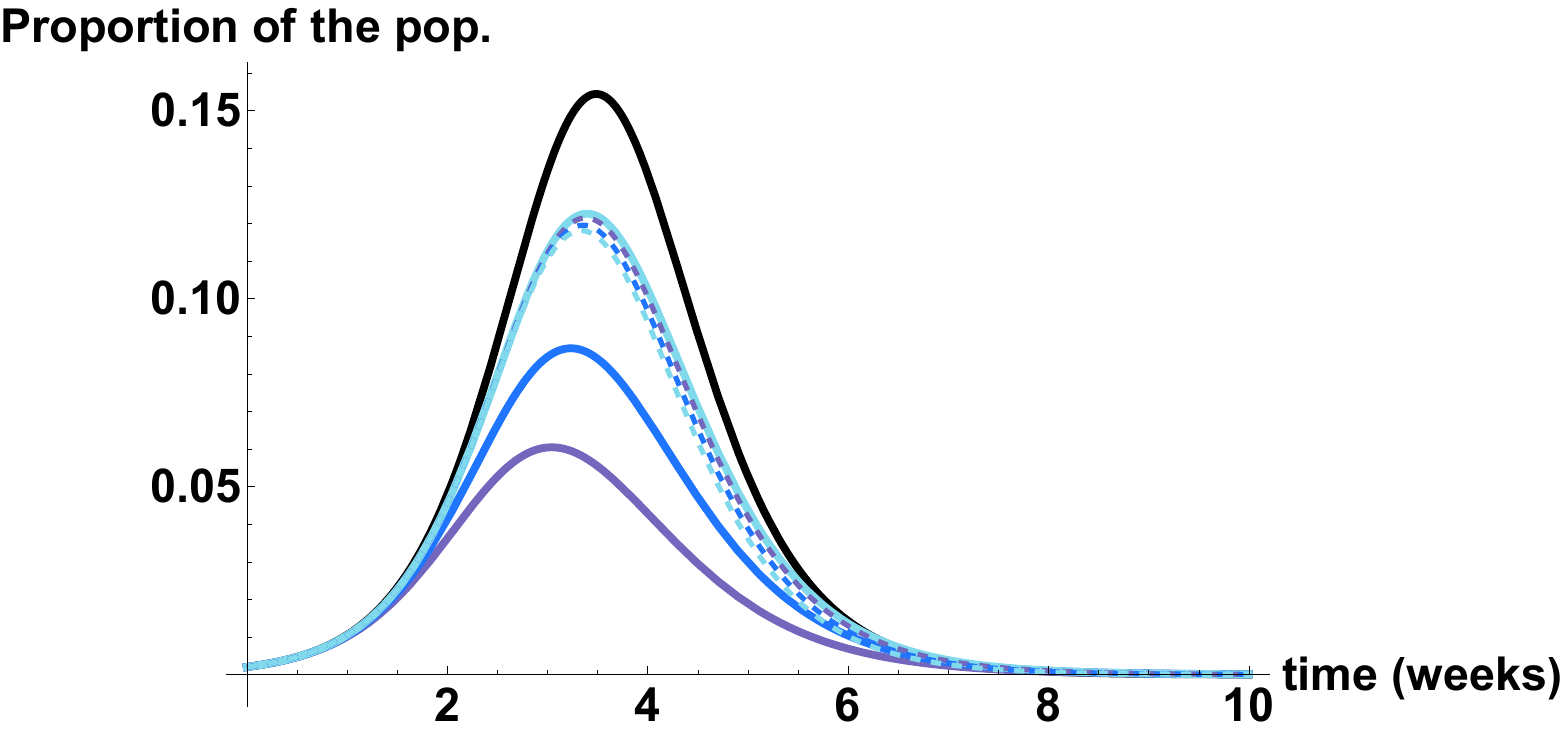}
     \includegraphics[width=0.45\textwidth]{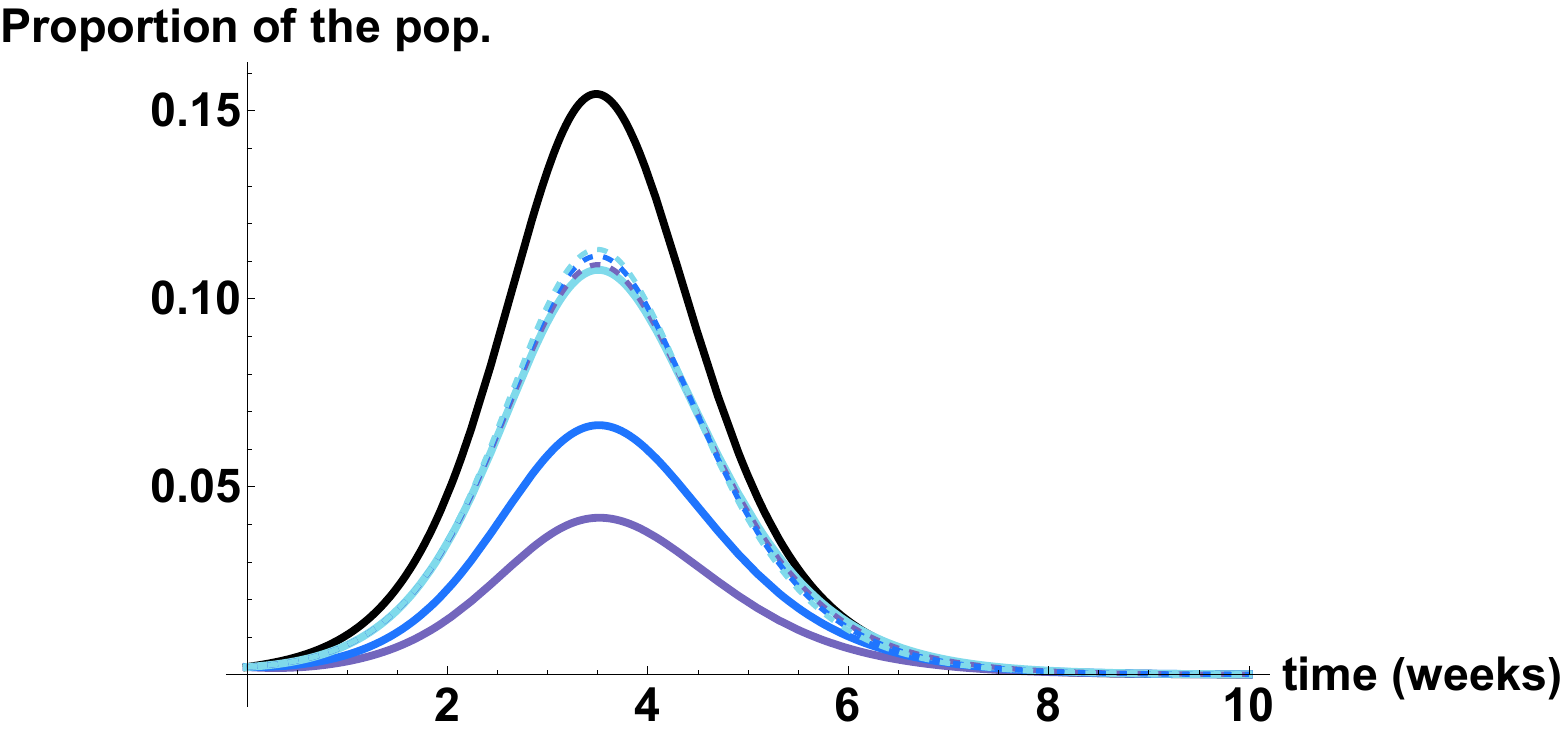}
    \includegraphics[width=0.45\textwidth]{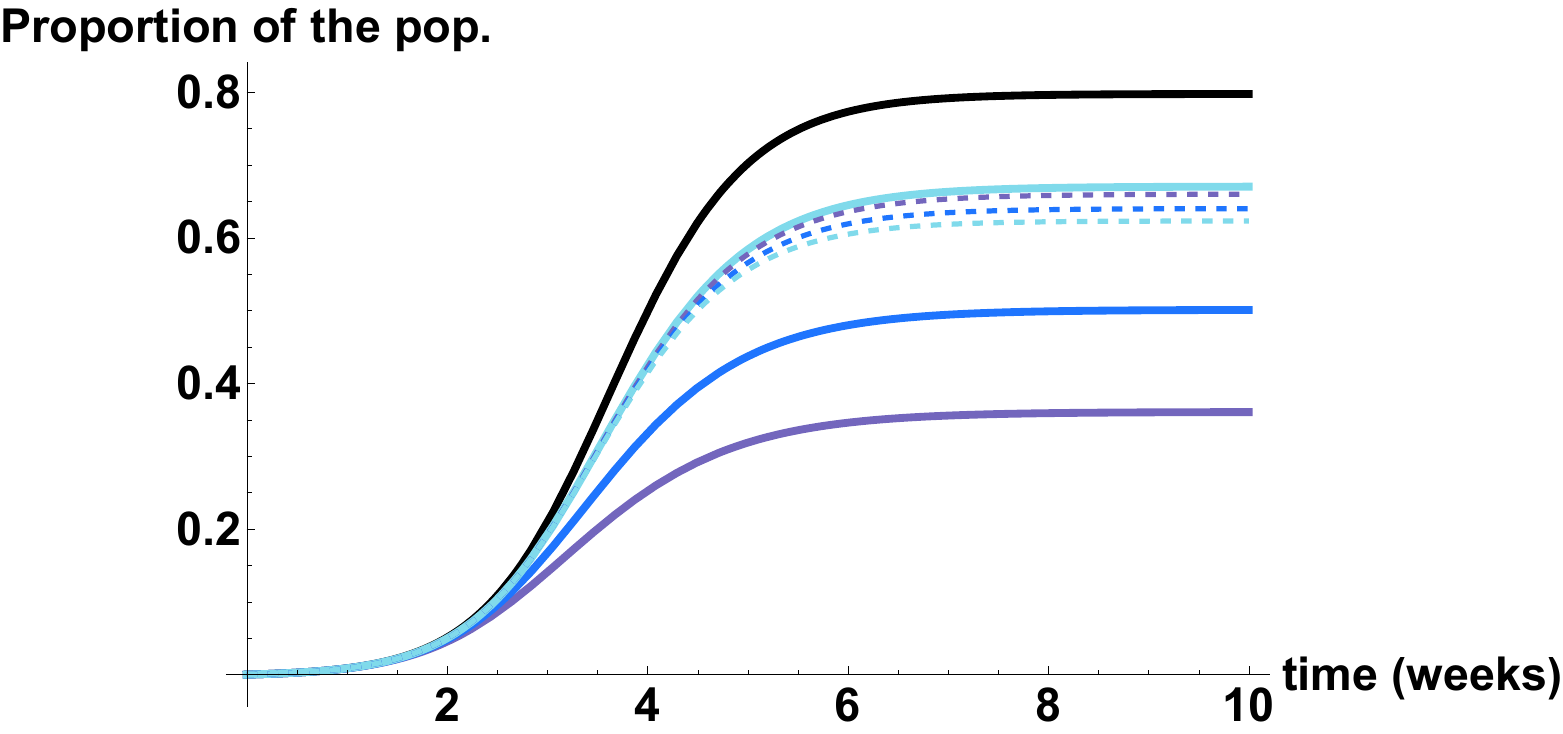}
    \includegraphics[width=0.45\textwidth]{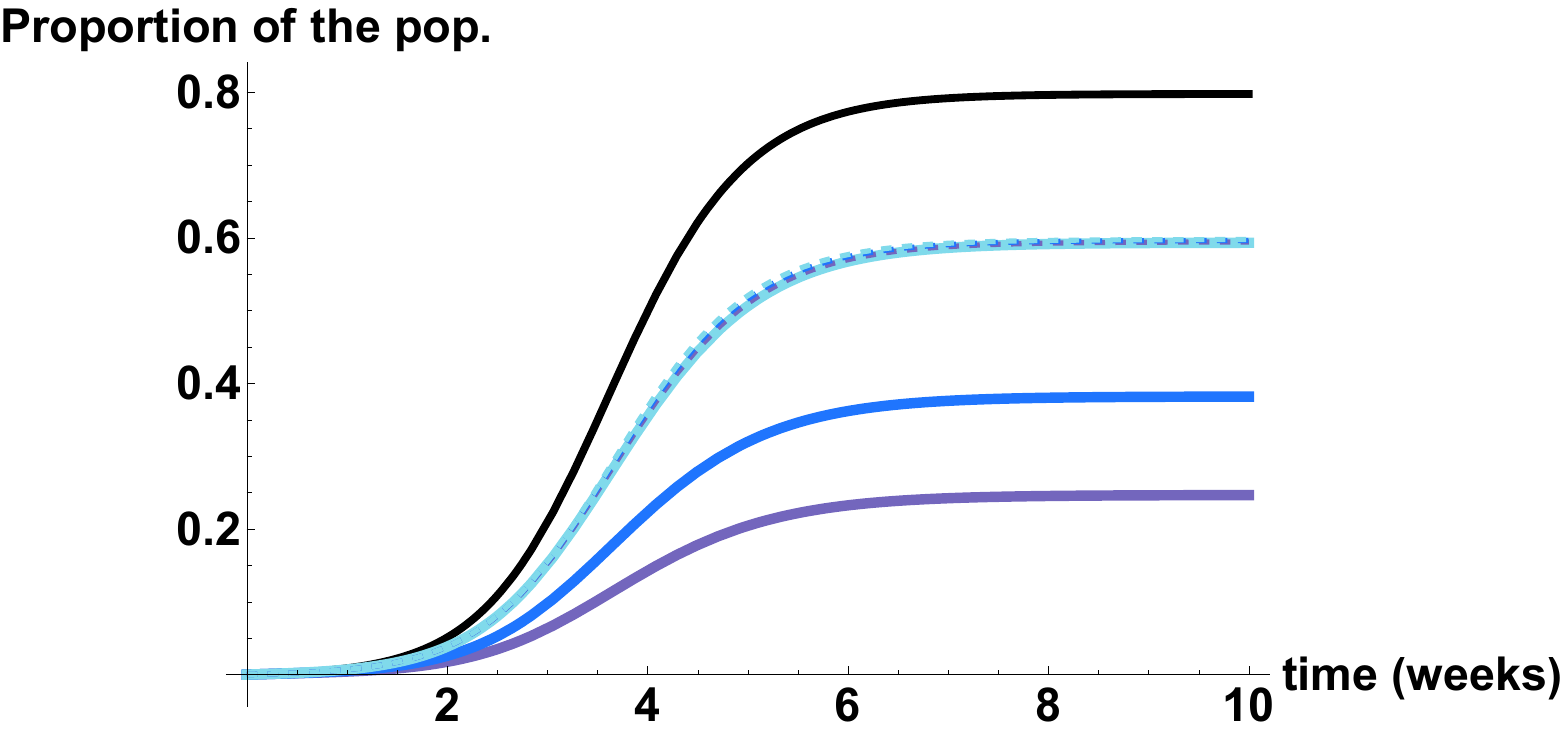}
    \includegraphics[width=\textwidth]{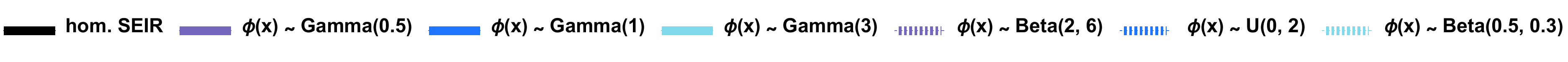}
    \caption{Plots showing proportion of $E(t) + I(t)$ (above) and $R(t)$ (below) for differently distributed susceptibility over time (all with  
$\mathcal{R}_0 = 2$). On the left panels, plots are shown with infectiousness independent of susceptibility; on the right panels, one can see plots with directly correlated infectiousness and susceptibility. The black curve represents the homogeneous SEIR model, coloured solid curves represent gamma distributions with different variance (see legend), coloured dashed curves represent Beta-distributed heterogeneity, with the same variance $(Var(X)=1/3)$ as the light-blue, solid curve. Note that $U(0,2)$ is the uniform distribution in $(0,2)$, which is the same as the extended Beta with $a=b=1$.}
    \label{fig1}
\end{figure}

%To illustrate the impact of heterogeneity in host susceptibility on epidemic outcomes, we present in Figure \ref{fig1} the simulations of the systems (precisely, the curves of prevalence and of recovered individuals, from which the final epidemic size can be assessed) using different Gamma and Beta susceptibility distributions. The distributions used (shown in Figure \ref{fig0}) are Gamma distributions with variance 1/3, 1, or 2, and three Beta distributions, all with variance 1/3.
To illustrate how heterogeneity in host susceptibility shapes epidemic dynamics, Figure \ref{fig1} presents simulations of the model systems. Specifically, we show the prevalence curves and the trajectories of recovered individuals, from which the final epidemic size can be inferred. The susceptibility distributions used in these simulations (displayed in Figure \ref{fig0}) include gamma distributions with variances of 1/3, 1, and 2, as well as three Beta distributions, each with variance 1/3.
All simulations shown in Figure \ref{fig1} have the same basic reproduction number 
$\mathcal{R}_0 = 2$ and include the homogeneous SEIR model (black lines), simulations of the models with independent susceptibility and infectiousness (left panels) or with correlated infectiousness (right panels); both panels include simulations where susceptibility is gamma-distributed so that either equations \eqref{sys1} or \eqref{sys2}--\eqref{sys2b},
and simulations with Beta-distributed susceptibility, in which case systems \eqref{sys_bar} or \eqref{sys_bar2}--\eqref{sys_bar2b} are used, together with \eqref{Phi_beta2}.

It can be seen that the homogeneous SEIR model yields the largest infection peak and the largest final size. Moreover, as the variance of the gamma distribution increases, the final size systematically decreases, both when susceptibility and infectiousness are independent and when they are correlated.  This dependence of the final size on the variance will be proved in the next section. One can also observe that different Beta distributions with the same variance produce very similar final epidemic sizes, both among themselves and relative to the gamma distribution with the same variance.
In this sense, the epidemic dynamics appear to depend primarily on the variance of the susceptibility distribution rather than on its specific shape: although the three distributions shown in the right panel of Figure \ref{fig0} differ markedly in form, they generate very similar trajectories and final epidemic sizes.

\section{Final size equation for the Gamma and Beta distributions}\label{Sec:finalsize}
\subsection{The case of independent infectiousness and susceptibility}
From the general formula \eqref{R0}, we know that when infectiousness and susceptibility are independent, $R_0 = \beta/\gamma$, regardless of $p$.
For $\Ro > 1$, the solution to equation \eqref{finalsize_gen} in $(0,1)$ gives $\bar s_\infty = \lim_{t\to+\infty} \bar s(t)$. In this situation, however, it is more convenient to work with an equation for $ S_\infty = \lim_{t\to+\infty} S(t)$. Although equation \eqref{finalsize_gen} could be reformulated in terms of $S_\infty$, it is simpler to obtain a prime integral directly from \eqref{sys1}.

In fact, it is easy to see from \eqref{sys1} that
\[
\frac{\dd}{\dd t}\left[ S(t)^{-1/p}+R_0(S(t)+E(t)+I(t))\right] = 0.
\]
Hence, one obtains
\begin{align} \label{fs}
    S_\infty^{-\tfrac{1}{p}} + \frac{R_0}{p} \cdot S_\infty = 1 + \frac{R_0}{p}.
\end{align}

% The following plots suggest that $S_\infty$ is decreasing in $p$ for different, appropriate $R_0$-s.
% \begin{figure}[H]
% 	\centering
% 	\includegraphics[width=0.3\textwidth]{./figures/R01.1.pdf}
%     \includegraphics[width=.3\textwidth]{./figures/R02.pdf}
%     \includegraphics[width=.3\textwidth]{./figures/R04.pdf}
%     \caption{$R_0 = 1.1,\, R_0 =2,\, R_0 =4$ from left to right.}
% \end{figure}

Concerning the solutions of the final size equation \eqref{fs}, we have the following 
\begin{theorem} If $\Ro > 1$, equation \eqref{fs} has a unique solution $S_\infty(p) \in (0,1)$. Moreover, $S_\infty(p)$ is a decreasing function of $p$.
\label{theor_ind}
\end{theorem}

Since the variance of the gamma distribution decreases with increasing $p$, this Theorem shows that the epidemic final size $(1-S_\infty(p)$ decreases with increasing variance. 
This result had been essentially proved by \pcite{Novozhilov2008} with completely different notation and methods.

\subsection{The case of correlated susceptibility and infectiousness}
When susceptibility and infectiousness are correlated, we start by obtaining a prime integral from equations \eqref{sys2}.
One finds that
$$
\frac{\gamma p}{\beta}S(t)^{-\tfrac{1}{p}} -\left( S_\infty^{1+\tfrac{1}{p}}  + U(t) + V(t)\right)
$$
is constant along the solutions to \eqref{sys2}.
Moreover, it is not difficult to show that necessarily 
$$ \lim_{t\to+\infty}U(t) = \lim_{t\to+\infty}V(t) = 0.$$
Then using $S(0) \approx 1$, $U(0) \approx V(0) \approx 0$, one obtains
\begin{align}
    \frac{p+1}{R_0}S_\infty^{-\tfrac{1}{p}} + S_\infty^{1+\tfrac{1}{p}} = 1 + \frac{p+1}{R_0}, \label{eqcorr}
\end{align}
where we have used the expression %$R_0=\dfrac{\beta}{\gamma}\left(1+\dfrac{1}{p}\right)$ (remember \eqref{R0}).
$R_0=\left(1+1/p \right)\beta/\gamma$ given by \eqref{R0}.

Concerning the solutions of the final size equation \eqref{eqcorr}, we have the following 
\begin{theorem} If $\Ro > 1$, equation \eqref{eqcorr} has a unique solution ${S^*_\infty}(p) \in (0,1)$.\\ Moreover, ${S^*_\infty}(p)$ is a decreasing function of $p$.\label{theor_corr}
\end{theorem}

Theorem \ref{theor_corr} shows, also for the case of correlated susceptibility and infectiousness, that the epidemic final size $(1-S^*_\infty(p))$ decreases with increasing variance.

\begin{figure}
    \centering
   \begin{tabular}{cc}
       $R_0=1.5$ & $R_0=3$ \\
     \includegraphics[width=0.45\linewidth]{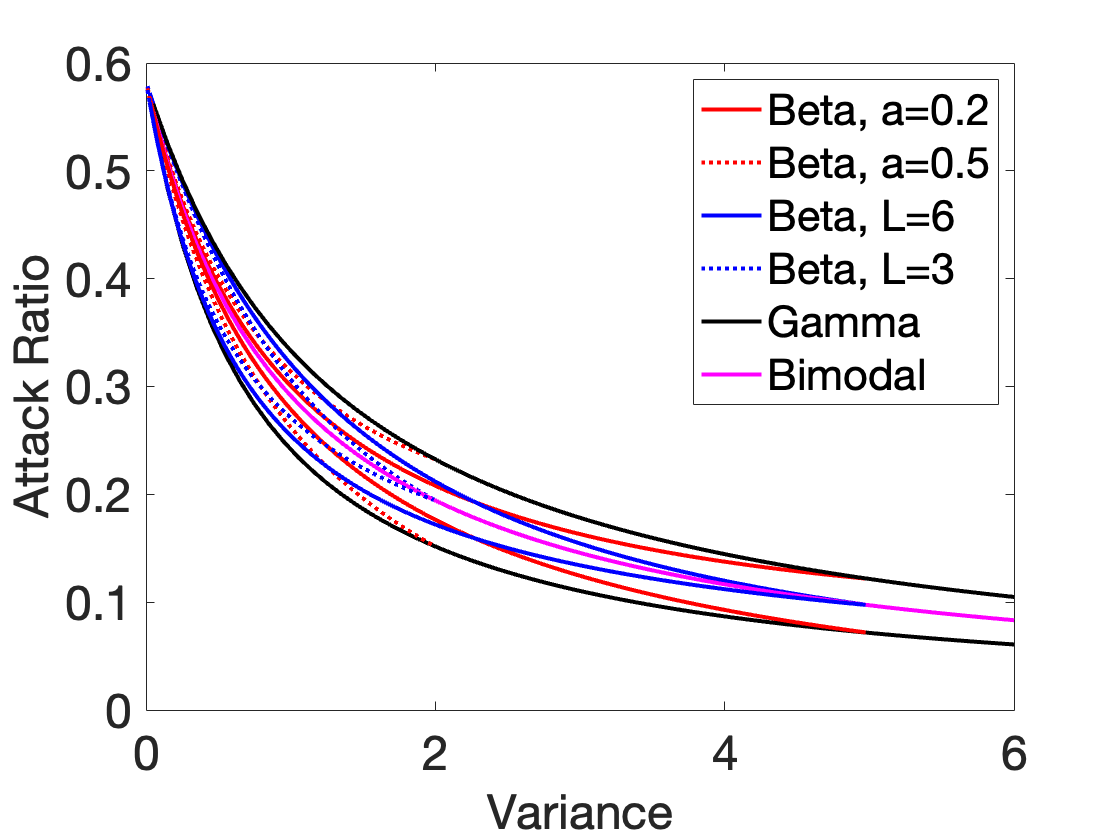}\ 
        & 
     \includegraphics[width=0.45\linewidth]{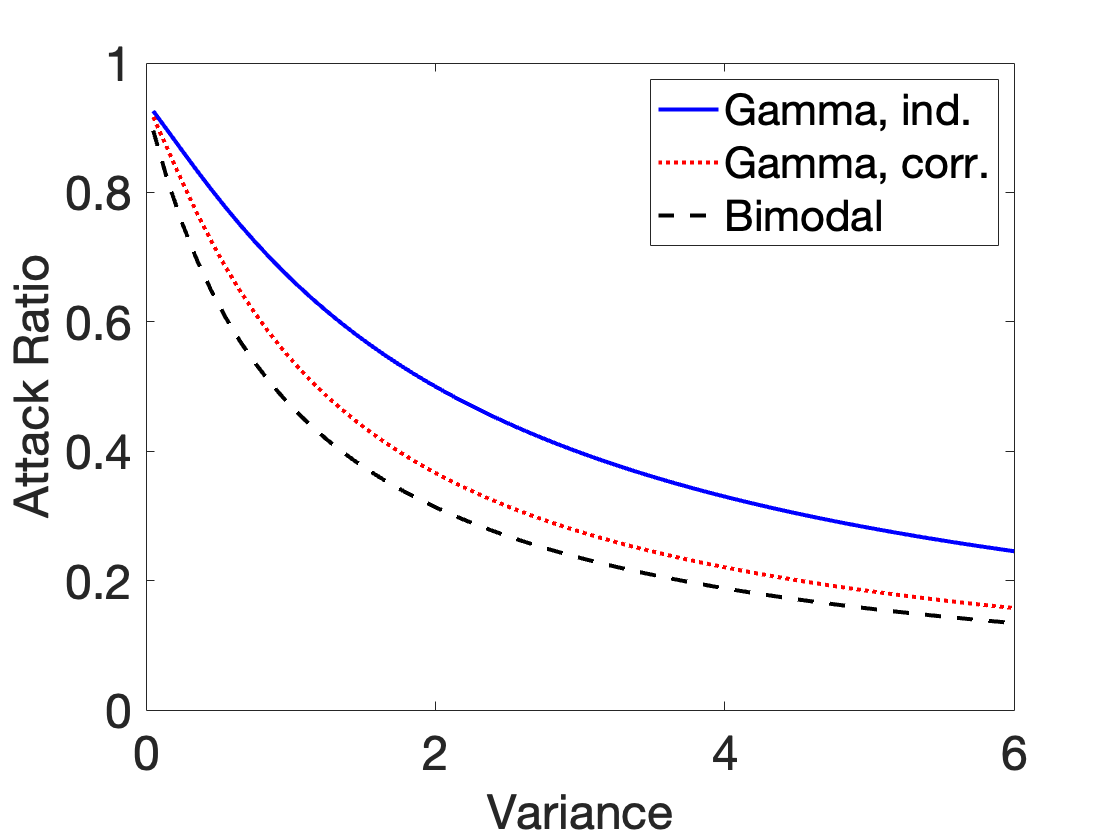}\ 
         \end{tabular}
   \caption{Left panel: Attack ratio at the end of an epidemic for systems \eqref{sys_bar} and  \eqref{sys_bar2}--\eqref{sys_bar2b}, plotted as a function of the variance with $R_0=1.5$; curves are shown when susceptibility $X$ follows the gamma distribution, or the Beta distributions with either the parameter choices in \eqref{choice_beta1} or \eqref{choice_beta2} and when $X$ is either equal to 0 or to $V+1$ (bimodal distribution). Each distribution is shown (with the same line type) both for independent (the higher curve) and for correlated (the lower curve) infectiousness. Right panel: Same with $R_0=3$; only the curves corresponding to the Gamma (both with independent anc correlated infectiousness) and bimodal distributions are shown.}
    \label{fig:finalsize_anal}
\end{figure}
Finally, we compare the final size in the two cases. We have the following 
\begin{proposition}
For the same values of $\Ro$ and $p$, let $S^*_\infty(p)$ be the solution to \eqref{eqcorr} and $S_\infty(p)$ the solution to \eqref{fs}. Then
$ S^*_\infty(p) > S_\infty(p) $.
\label{prop_compare}
\end{proposition}
In words, for the same value of $\Ro$, the susceptible fraction at the end of an epidemic outbreak is higher when susceptibility and infectiousness are correlated than when they are independent; hence, the attack ratio is lower with correlation between susceptibility and infectiousness.

We considered the final size also when susceptibility follows the (extended) Beta distribution, but this was possible only numerically, as shown below. We start by considering the limiting case, introduced in the previous Section, of a bimodal distribution, in which susceptibility is either equal to 0 (with probability $(L-1)/L$) or to $L$ (with probability $1/L$).
Note that the probabilities have been set to have average susceptibility equal to $1$ and that their variance is equal to $L-1$. 
\\
This case actually corresponds to the standard epidemic model, where only a fraction $1/L$ is initially susceptible (while the rest of the population is immune) and one would set $S_0=1/L$, and the contact parameter would be defined as $\tilde \beta = \beta L$, and the basic reproduction number $\tilde \Ro = \tilde \beta/ \gamma$, while what here is defined as $\Ro = \beta/\gamma$ would be the effective reproduction number.
One can then use the standard formula for $\bar x = S_\infty/S_0$ that is obtained (if $\Ro = \tilde \Ro S_0 > 1$ as the only solution in $(0,1)$ of
\begin{equation}
    \label{FKMK}
    F(x) := \log(x)+\Ro(1-x) = 0.
\end{equation}
Hence, the attack ratio is $S_0 (1-\bar x )= \dfrac{1 - \bar x}{V+1}$.
\\
Note, finally, that in this case one cannot distinguish between independent and correlated infectiousness, since only those who are susceptible can be infected.
\\
Comparing this case with the case where the susceptibility follows a Gamma-distribution, we have the following
\begin{proposition}
\label{prop_bimodal_gamma}
$\dfrac{1 - \bar x}{V+1} < 1 - S_\infty(p)$ where $\bar x$ is solution of \eqref{FKMK}, $S_\infty(p) $ solution of \eqref{fs}, and $p =1/V$.
\end{proposition}
   In words, the attack ratio found when susceptibility is bimodal is lower than that obtained when susceptibility is gamma-distributed with the same variance and the same value of $\Ro$.
  \\ 
We complete the analysis with numerical computations of the attack ratio. In Figure \ref{fig:finalsize_anal}, we show how the final attack ratio (the fraction of the population infected during an epidemic outbreak) depends on the variance for some distributions of susceptibility. In the left panel ($\Ro=1.5$), we show the results for the gamma distribution and for the Beta distribution under the parametrisations \eqref{choice_beta1} and \eqref{choice_beta2} with either independent or correlated infectiousness, and also for the bimodal distribution.  The panel, beyond numerically confirming the results of Theorems \ref{theor_ind} and \ref{theor_corr}, of Propositions \ref{prop_compare} and \ref{prop_bimodal_gamma}, shows that the attack ratios obtained with the Beta distributions are intermediate between those with the gamma distribution and the bimodal case for the same value of the variance and the same type of infectiousness (independent or correlated to susceptibility). One can also see that, for the same distribution of susceptibility, the attack ratio when infectiousness is correlated to susceptibility is lower than when they are independent, extending the result proved for the gamma distribution in Proposition \ref{prop_compare}. Furthermore, in the limiting cases where the Beta distribution tends either to the Gamma or to the bimodal case, the attack ratio converges, as expected, to the corresponding values. Finally, one sees that, for all distributions, the attack ratios obtained when infectiousness is correlated to susceptibility are lower than for the bimodal case.
\\
In the right panel, we show how the final attack ratio depends on the variance of susceptibility when $\Ro=3$, only for the gamma distribution (both with independent and correlated infectiousness) and for the bimodal case. It can be seen that the final observation made when $\Ro=1.5$ does not hold: in this case, the attack ratio with bimodal susceptibility is lower than with correlated infectiousness. All other observations concerning Beta distributions (the attack ratios being intermediate between the corresponding Gamma and bimodal, converging to the expected limiting case, being lower with correlated than with independent infectiousness) are instead confirmed. This can be seen in the Figures shown in the Supplementary Material (Section~\ref{sec:suppl_fig}), since we avoided showing the corresponding lines in the right panel, which would have become very messy with all the lines.

Overall, one can say that the final epidemic size obtained from the Beta distribution is close to that produced by a gamma distribution with the same variance and the same choice of either independent or correlated infectiousness, though some differences can be seen. %\deleted{In particular, the Beta distributions yield slightly lower attack ratios than the Gamma when susceptibility and infectiousness are independent, but slightly higher when they are correlated.}  
When comparing Beta and gamma distributions, or Beta distributions with different parametrisations, it is possible to find pairs of distributions for which the one with the lower variance also has a lower attack ratio, in line with the observation of \tcite{Novozhilov2008} that mean and variance do not fully characterise the behaviour of heterogeneous models.  Nevertheless, within each fixed Beta parametrisation, increasing the variance consistently leads to a lower attack ratio.

%\textcolor{black}

\section{Model fit to data on seasonal influenza in Italy}\label{sec:datainfluenza}
\subsection{Data description}
The data used in this study were obtained from RespiVirNet \pcite{ISS}, which provides weekly, age-stratified influenza-like illness (ILI) incidence rates from the national sentinel surveillance system. Raw records include ILI aggregated case counts (and number of patients in the sentinel cohorts) from the 2014-15 season onwards, reported by age groups (0--4, 5--14, 15--64, 65 and above) and week (from Week 46 to Week 17 of the subsequent year). Furthermore, data include, for each week, the fraction of virological samples from the sentinel network that are positive to the different influenza strains; these are age-stratified only in recent years, thus, for consistency, we did not use this information.

To obtain an estimate of the incidence of influenza infections in week $w$ of year $y$, we corrected the raw data by the following formula  
\begin{equation}
    \label{casi}
    \textit{cases}_{y, w} = \sum_{\textit{a $\in$ age}}\textit{ILI}_{y, w, a} \cdot \frac{{N}_{a} }{1000}\cdot \textit{pos}_{y, w} \cdot ur_{a},
\end{equation}
where 
\begin{itemize}
    \item $\textit{ILI}_{y, w, a}$ is the reported incidence rate (per 1,000) of ILIs in age-group $a$ in week $w$ of year $y$,
    \item ${N}_{a}$ is the number of individuals in age group $a$,
    \item $\textit{pos}_{y, w}$ is the ratio of samples positive to influenza among sentinel swabs in week $w$ of year $y$,
    \item $ur_a$ is the estimate of age-dependent underreporting ratio \pcite{Trentini2022}.
\end{itemize}
The influenza seasons considered range from 2014-15 to 2023-24. However, the seasons 2020-21, 2021-22, and 2022-23 have been excluded from the analysis since the lockdowns and other protective measures strongly affected the dynamics (no influenza cases were recorded in 2020-21, very few in 2021-22, while 2022-23 had an anomalous pattern, incompatible with a SEIR model with constant parameters).

\subsection{Fitting methods}
We calibrated homogeneous and heterogeneous gamma-distributed SEIR-based models to describe the transmission dynamics, fitting the estimates obtained of weekly influenza infections in each given epidemic season. 

For the model \eqref{sys1} with gamma-distributed susceptibility, we considered as unknown parameters $\beta$, $p$, and the initial condition $I_0$ assuming for simplicity that $E(0)=I(0)=I_0$, where $0$ is the first week of each epidemic season; at $t=0$ everybody else was assumed to be susceptible, as nobody was considered completely immune to the circulating strain at the season start. In the case of the homogeneous SEIR model (formally, it can be seen as model \eqref{sys1} with $p=\infty$), we instead fitted $\beta$ and the initial conditions $E(0)=I(0)$ and $R(0)$. For both models, we set $\alpha=\gamma = 4$ (weeks$^{-1}$), corresponding to an average length of generation time of 3.5 days \pcite{Vink2014}.

The parameters were estimated through non-linear least squares using the function \texttt{scipy.optimize\_fit}. 
For each model variant, global minima were identified by performing a grid search: the fitting routine was evaluated across the grid, and the parameter set yielding the highest $R^2$-score (coefficient of determination) was retained. %In order to identify the global minima, for each model variant we implemented grid search procedures by looping over the grid, applying the fitting routine, and eventually retaining the parameter set that maximizes the $R^2$-score (coefficient of determination).

The final output of each procedure includes the best-fitting parameters, the corresponding fitted trajectory, and the maximum $R^2$ value achieved. These outputs have been used to compare the performance of the two models on the same data. Since we aimed to test whether these simple models, with and without heterogeneity, could reasonably describe the available data, we did not perform rigorous statistical model comparisons.

\subsection{Results}
The results of the fitting procedure are summarised in Tables \ref{tab:SEIR_hom} for the homogeneous SEIR model and in Table \ref{tab:SEIR_het} for the SEIR model with gamma-distributed susceptibility. For both cases, the maximum $R^2$ value achieved is reported. Beyond the value of $R^2$, we also include in the tables the estimated effective reproduction numbers $\mathcal{R}_0$ ($\beta/\gamma \cdot R(0)/N$ for the homogeneous model, $\beta/\gamma$ for the one with heterogeneity), the initial immune fraction $R(0)/N$ for the homogeneous model, $p$ (the parameter of the gamma distribution) for the model with heterogeneity.
%The influenza seasons considered range from 2014-15 to 2023-24. However, the seasons 2020-21, 2021-22 and 2022-23 have been excluded from the analysis, since the lock-downs and other protective measures had strongly affected the dynamics (no influenza cases have been recorded in 2020-21, very few in 2021-22, while 2022-23 had an anomalous pattern, incompatible with a SEIR model with constant parameters).

Across all seasons, both models achieve $R^2$ values close to 1, indicating an excellent fit to the data, with only small distinctions in their performance.
%One can see that both models yield an extremely good fit to the data with $R^2$ very close to 1 in all seasons, and minor differences between the two models.

One may also observe that the estimated values of $\mathcal{R}_0$ are very similar between the two models (though consistently larger for the model with heterogeneity) and quite larger than the corresponding estimates by \tcite{Trentini2022} for the same years.

Furthermore, in the homogeneous model, the initial immune fraction is estimated to be quite large (70\% or above in all seasons except 2017-18). Correspondingly, in the model with heterogeneity, the variance in susceptibility is estimated to be very large (remember that the variance is equal to $1/p$) with a minimum in 2017-18 season.

\begin{figure}
\centering
\includegraphics[width=0.49\textwidth]{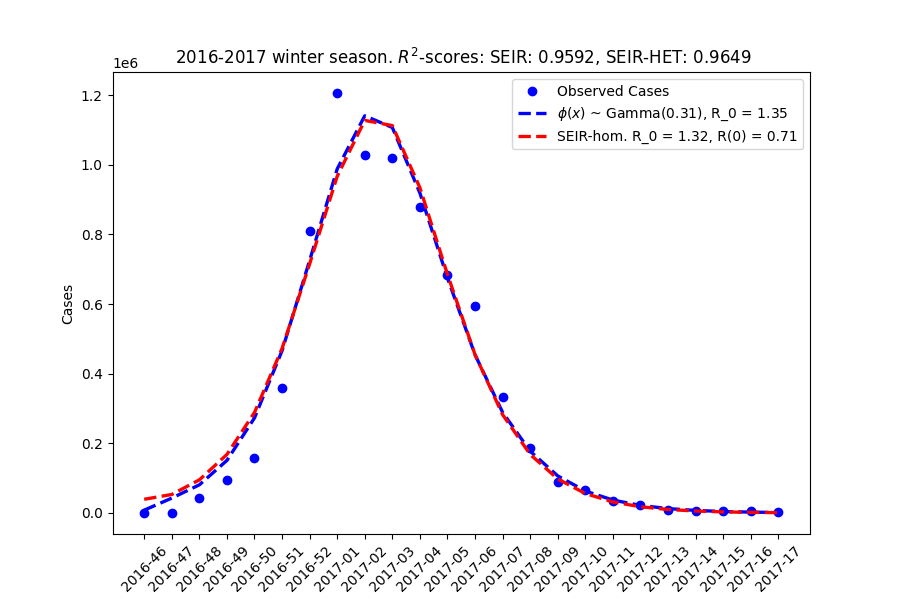}
    \includegraphics[width=.49\textwidth]{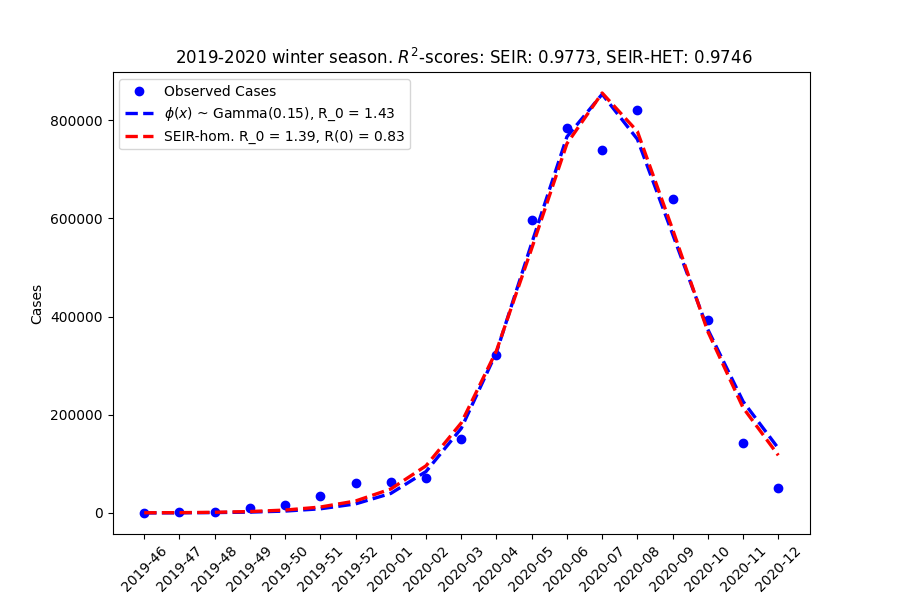}
    \\
\includegraphics[width=0.49\textwidth]{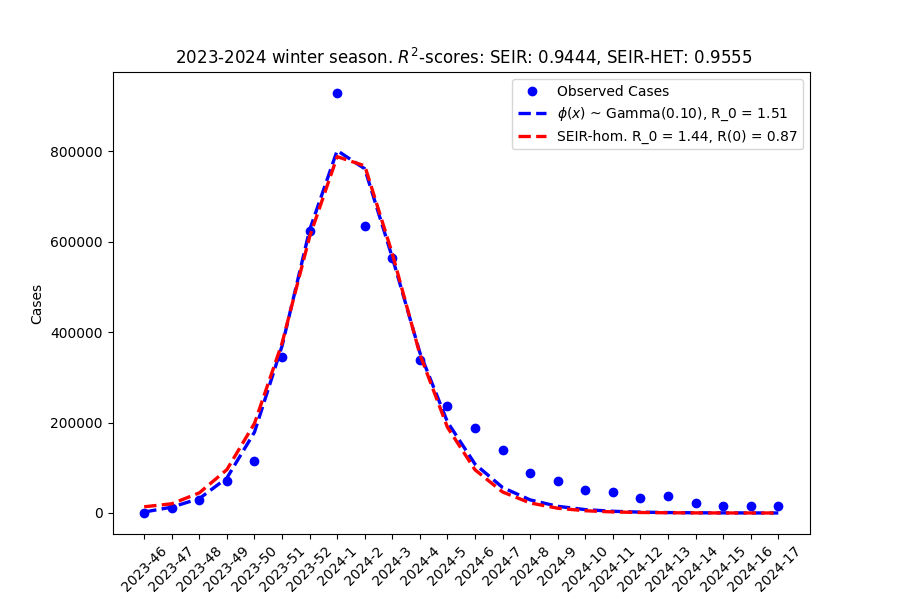}
    \caption{Comparisons between the weekly infections estimated from the homogeneous or heterogeneous SEIR models (dashed lines) and the estimated cases from the data using formula~\eqref{casi} (dots) in the seasons 2016-17, 2019-2020, and 2023-24}
\label{fig:fit}
\end{figure}

%In Figure \ref{fig:fit}, we show the resulting fit of the two models for selected seasons: precisely 2016-17 and 2023-24 are the two seasons in which the difference in $R^2$ between the two models is highest, while 2019-20 is the only season in which the $R^2$ with the homogeneous model is higher than with the heterogeneous model.
Figure \ref{fig:fit} illustrates the fitted curves for selected seasons. We highlight the 2016–17 and 2023–24 seasons because these are the years for which the $R^2$ values of the two models differ the most. In contrast, 2019–20 stands out as the only season where the homogeneous model achieves a higher $R^2$ than the heterogeneous one.

\begin{table}
\begin{subtable}[b]{1\textwidth}
\centering
{\begin{tabular}{ |c|c|c|c|c|c|c|c|} 
\toprule\\[-0.65cm]
 & \textbf{2014} & \textbf{2015} & \textbf{2016} & \textbf{2017} & \textbf{2018} & \textbf{2019} & \textbf{2023}\\
 \hline
$ R^2$ & 0.9731 & 0.9768 & 0.9592  & 0.9637 & 0.9929 & 0.9773 & 0.9444\\ 
\hline
 $R(0)/N$ & 0.69 & 0.76  & 0.71 & 0.46 & 0.77 & 0.83 & 0.87\\ 
 \hline
 $\mathcal{R}_0$ & 1.33 & 1.24  & 1.32 & 1.31 & 1.37 & 1.39 & 1.44\\[-0.08cm] 
 \bottomrule
\end{tabular} }   
\caption{Homogeneous SEIR model }
\label{tab:SEIR_hom}
\end{subtable}

\vspace{0.5cm}

\begin{subtable}[b]{1\textwidth}
\centering
 \begin{tabular}{ |c|c|c|c|c|c|c|c| } 
 \toprule\\[-0.65cm]
 & \textbf{2014} & \textbf{2015} & \textbf{2016} & \textbf{2017} & \textbf{2018} & \textbf{2019} & \textbf{2023}\\
 \hline
$ R^2$ & 0.9791 & 0.9776 & 0.9649  & 0.9686 & 0.9954 & 0.9746 & 0.9555\\ \hline
 $\textcolor{white}{ppp}p\textcolor{white}{ppp}$ & 0.33 & 0.26  & 0.31 & 0.85 & 0.21 & 0.15 & 0.1\\ 
 \hline
 $\mathcal{R}_0$ & 1.36 & 1.26  & 1.35 & 1.34 & 1.41 & 1.43 & 1.51\\[-0.08cm] 
 \bottomrule
\end{tabular}    
\caption{SEIR model with gamma-distributed susceptibility}
\label{tab:SEIR_het}
\end{subtable}
\caption{Main outputs of the fitting procedure for the homogeneous SEIR model (a) and for the SEIR model with gamma-distributed susceptibility (b).}
\end{table}

\begin{comment}
SEIR-het., $k = 1, \phi_X(x)$ $\sim$ Beta$(a, b, 0, L = 1 + \frac{b}{a})$
\begin{center}
\begin{tabular}{ c|c|c|c|c|c|c|c } 
 \hline
 & 2014 & 2015 & 2016 & 2017 & 2018 & 2019 & 2023\\
 \hline
$ R^2$ & 0.9784 & 0.9775 & 0.9637  & 0.9680 & 0.9952 & 0.9772 & 0.9546\\ \hline
 $\{a, b\}$ & \{0.3, 10\} &  \{0.24, 10\}  & \{0.28, 10\} & \{0.76, 10\} & \{0.20, 10\} & \{0.01, 0.05\} & \{0.1, 10\}\\ 
 \hline
 $\mathcal{R}_0$ & 1.35 & 1.26  & 1.35 & 1.33 & 1.41 & 1.39 & 1.50\\  \hline
  $Var(X)$ & 2.91 & 3.74  & 3.12 & 1.12 & 4.47 & 4.95 & 9.28\\  \hline
\end{tabular}
\end{center}
\end{comment}

\section{Discussion}\label{sec:conclusion}
The paper focuses on incorporating host heterogeneity in susceptibility and infectiousness into SEIR epidemiological models. The method used is based on the results by \tcite{DiekmannInaba2023}, which have been adapted specifically for the SEIR model; we believe that the presentation for this specific setting may be easier to follow than the general treatment in \pcite{DiekmannInaba2023}. 

Following this path, in the case of a generic distribution of susceptibility, one obtains system \eqref{sys_bar2} when susceptibility and infectiousness are independent, whereas system \eqref{sys_bar2b} when they are correlated. The cases of a gamma distribution (already considered by \tcite{DiekmannInaba2023}) and of a Beta distribution of susceptibility are then illustrated.

By studying the equation for the final size of the susceptible fraction, we show that, for any distribution of susceptibility, the final size of an epidemic (i.e., the total fraction of infected individuals) heterogeneity reduces the epidemic final size compared to homogeneous models with the same value of the reproduction ratio $\Ro$. This result holds in both cases where susceptibility and infectiousness are independent and when they are correlated.

The result is consistent with the findings by \tcite{Novozhilov2008}, which followed a very different approach and did not consider the case of correlated susceptibility and infectiousness.

In the case where susceptibility is gamma-distributed, we also show that the final size decreases as a function of the variance of the gamma distribution, both with independent and correlated infectiousness. Furthermore, the final size is smaller with correlated than with independent infectiousness for the same variance of the distribution. In biological terms, greater variability in the host population dampens epidemic spread, as the more resistant individuals limit chains of transmission. 
Notice that assuming that infectiousness is proportional to susceptibility may cover the case in which the difference in susceptibility reflects differences in behaviour rather than in immunological response to infection challenges. In such a case, in fact, individuals with more contacts, will be, on the one hand, more likely to get infected, and, on the other hand, more infectious, assuming that differences in behaviour are retained also after infection.

Empirical fits to Italian influenza surveillance data indicate that the model with gamma-distributed heterogeneity in susceptibility is sufficiently flexible to adapt to the dynamics of all normal seasons of influenza infections. 
Indeed, the homogeneous SEIR model can also accurately describe the dynamics of seasonal influenza, but only assuming that a considerable fraction of individuals were already immune to the circulating strain at season start; however, estimates, based on serological data, of the fraction of the population immune to the circulating strains at season start \pcite{Lunelli2013} are much lower than the estimates for $R(0)$ in Table \ref{tab:SEIR_hom}. 
Thus, in order to fit the available data, the homogeneous SEIR model must make unrealistic assumptions about pre-existing immunity.

Notice that we only considered retrospective fits to the infection data over a whole season. Predicting the time course of an influenza season is notoriously an extremely different task to which many different methods have been applied, with mixed success \pcite{Viboud2019,Reich2019}.

We believe, rather, that epidemic outbreak models should incorporate not only known sources of heterogeneity, such as age, spatial location, or social class, but also unobserved variation in susceptibility, represented by an additional parameter (such as the $p$ used here).
Such intrinsic heterogeneity may reflect not only biological differences among individuals but also differences in behaviour that affect the probability of getting infected, as in \pcite{BrittonBallTrapman2020}. 
Heterogeneity may also be tied to the evolution of strains, so that individuals may have been last infected by strains with different similarities to the currently circulating strain \pcite{Andreasen2003a, Britton2025}.

From a public health perspective, these findings underscore the importance of incorporating host heterogeneity into epidemic forecasting and the design of intervention strategies. Neglecting such variability may result in inaccurate estimates of epidemic size and misidentification of critical thresholds. At the same time, our models rely on simplifying assumptions, most notably homogeneous mixing and fixed susceptibility distributions, which should be kept in mind when interpreting the results.

It is easy to extend the current model to allow for multiple groups connected through a given contact matrix.
From a practical point of view, the question is whether one can assume the same distribution of susceptibility for each group or whether it would be necessary to allow for different distributions, leading to a multiplication of parameters. We plan to explore this issue by extending the present analysis to include age structure \pcite{Trentini2022} in the model and in the available data, as well as known vaccination rates and effectiveness \pcite{Calabro2025}.

Other extensions, such as including temporal changes in susceptibility (e.g., due to waning immunity), seem more complex. 
More generally, the model is tailored for a single outbreak, where it is assumed that 
complete immunity is absent at the start, and susceptibility is a static parameter.
Formula \eqref{s(t,x)}, which was essential in the analysis, could be extended to
$$ s(t,x) = s(0,x) \left(\frac{\bar s(t)}{\bar s(0)}\right)^x,$$
if all $s(0,x)$ were known. However, relating $s(0,x)$ to the infection dynamics in previous seasons would be rather complicated, especially if the intrinsic parameter $x$ were allowed to vary.

Despite these difficulties, we believe that our results reinforce the central role of heterogeneity in shaping epidemic trajectories during an outbreak and provide a feasible modelling framework for influenza and other respiratory infections.\\[0.3cm]

\noindent
\textbf{Acknowledgments.}  This work was supported by the project ``One Health Basic and Translational Actions Addressing
Unmet Needs on Emerging Infectious Diseases'' (INF-ACT), BaC ``Behaviour and sentiment monitoring and modelling for outbreak control/BEHAVE-MOD'' (No. PE00000007, CUP I83C22001810007)
funded by the NextGenerationEU.
TT was also supported by PRIN 2020 project No. 2020JLWP23 and by RRF-2.3.1-21-2022-00006.
AP and CS are members of the \emph{Unione Matematica Italiana} (UMI) group ``\emph{Modellistica Socio-Epidemiologica}'' (UMI-MSE). AP is a member of INdAM-GNAMPA Group, CS of the INdAM-GNFM Group. The authors thank Andrea Bizzotto for the fruitful discussion about prediction of influenza epidemics. 

\printbibliography

\appendix
\renewcommand{\thesection}{\Alph{section}} % corrected redefinition of '\thesection'
\makeatletter
\renewcommand\@seccntformat[1]{\appendixname\ \csname the#1\endcsname.\hspace{0.5em}}
\makeatother

\section{Proofs}\label{sec:proofs_A}
In the following, the proofs of the theorems stated in the paper can be found.
\begin{proof}[Proof of Theorem \ref{theo3.1}]
Let $G_k(w)= \Phi_{k-1}(w) - \gamma/\beta\log(w)$ for $k=1$ or 2. Clearly $\lim\limits_{w \to 0^+} G_k(w)=+\infty$, while $G_k(1)=1$ for $k=1$ or 2. Furthermore
$$ G_k'(w) = \frac{\Phi_k(w)- \gamma/\beta}{ w},\qquad G''_k(w) = \frac{\Phi_{k+1}(w)}{w^2 } - \frac{\Phi_k(w)- \gamma/\beta}{ w^2}. $$
These formulae show that $G_k'(\bar w) = 0 \Longrightarrow G''_k(\bar w) > 0$; hence, $G$ may have minima, but cannot have maxima. Finally we compute
 $$G'_k(1)=\Phi_k(1)- \frac{\gamma}{\beta}= \Phi_k(1)\left(1- \frac{1}{\Ro}\right),\qquad k=1\mbox{ or } 2.$$
Therefore, $G'_k(1) < 0$ if and only if $\Ro > 1$. Hence, equation \eqref{finalsize_gen} has a unique solution in $(0,1)$ if $\Ro > 1$, and no solution if $\Ro \le 1$.

Concerning the second claim, it is well known that the final size $ S_\infty$ of the homogeneous model satisfies the equation
\begin{equation}
\label{finalsize_hom}
 S_\infty - \frac{1}{\Ro}\log( S_\infty) =S_0 -  \frac{1}{\Ro}\log(\bar S_0) = 1.
\end{equation}
In case $k=1$, equation \eqref{finalsize_gen} reads $\Phi_{0}(\bar s_\infty) - 1/{\Ro}\log(\bar s_\infty) =1$. Since the function $w^x$ is a strictly convex function of $x$, Jensen's inequality implies 
$$ \Phi_0(w) = \int_0^\infty w^x \phi(x)\, \dd x = \E(w^X) > w^{\E(X)} = w, $$
where $X$ is a random variable with density $\phi(x)$. From the previous inequality, it follows that, if $k=1$, the LHS of \eqref{finalsize_gen} is larger than the LHS of \eqref{finalsize_hom} for any $w \in (0,1)$. This implies that $\bar s_\infty$, solution to \eqref{finalsize_gen} for $k=1$, is larger than $S_\infty$, solution to \eqref{finalsize_hom}. 
Furthermore, $\bar S_\infty = \Phi_0(\bar s_\infty) = \E(\bar s_\infty^X) >\bar s_\infty $, using again Jensen's inequality.

If $k = 2$, equation \eqref{finalsize_gen} reads $$\Phi_1(\bar{s}_{\infty}) - \frac{\Phi_2(1)}{\mathcal{R}_0} \log (\bar{s}_{\infty}) = 1. $$
Now consider $$H(w):=\frac{\Phi_1(w)}{\Phi_2(1)}-\frac{ \log (w)}{R_0} - \frac{1}{\Phi_2(1)} $$ and $$h(w):=w - \frac{ \log (w)}{R_0} - 1.$$ From what seen, we have $H(\bar s_\infty)=h(S_\infty)=0$. We now prove that $H(w) \geq h(w) $ on $(0,1)$ with equality if and only if $X \equiv 1$.
Equivalently, we prove that
$$I(w) := \Phi_2(1)(H(w)-h(w)) = \Phi_1(w)- w \Phi_2(1) + \Phi_2(1) -1 \geq 0 .$$
Using $\Phi_1(w)=\mathbb{E}[X w^X], \, \Phi_2(1) = \mathbb{E}[X^2]=Var(X)+1$, $\E(X)=1$, we have
\begin{equation*}I(w) = \mathbb{E}[X w^X] - w \mathbb{E}[X^2] + \mathbb{E}[X^2] -1. \end{equation*} 
Since the function $w^x$ is convex and $X \ge 0$, we have
\begin{equation*}
\begin{split}
    w^X \ge& w + w\log(w)(X-1) \\ &\Longrightarrow \mathbb{E}[X w^X] \ge w\E(X) + w\log(w) \E(X(X-1)) \\
    &\hspace{2.5cm}= w + w\log(w)Var(X).
    \end{split}
\end{equation*}
Then
\begin{equation*}
\begin{split}
I(w) &\ge w + w\log(w)Var(X) + \left(Var(X)+1\right)(1-w) -1 \\
&=  Var(X)(w\log(w) +1-w) > 0.
    \end{split}
\end{equation*}
In fact, if we let $q(w) = w\log(w) +1-w$, we have $q(1) = 0$, $q'(w)=\log(w) < 0$ for $w \in(0,1)$, which implies $q(w) > 0$ for $w \in(0,1)$.

From $H(w) > h(w)$ for $w \in(0,1)$, it follows  $\bar s_\infty > S_\infty$. As in the case of $k=1$, we have $\bar S_\infty > \bar s_\infty$, which completes the proof.
\end{proof}

\medskip
\begin{proof}[Proof of Theorem \ref{theor_ind}]
Consider the function
\begin{equation}
    \label{G1p}
    G(x,p) =  x^{-\tfrac{1}{p}} -1 - \frac{R_0}{p} (1-x).
\end{equation}
Obviously, a solution $S_\infty(p)$ to \eqref{fs} satisfies $G(S_\infty(p),p)=0$.

In general, we have
$$
G_x(x,p) = \frac1p \left(-x^{-\tfrac{1}{p}-1} +\Ro\right)\qquad
G_{xx}(x,p) = \frac1p\left(\frac1p+1\right)x^{-\tfrac{1}{p}-2}>0.
$$
Hence $G$, as a function of $x$, is convex. Furthermore
$$
\lim_{x \to 0^+} G(x,p) = +\infty,\qquad G(1,p) = 0,\qquad G_x(1,p) =\frac1p(\Ro-1).
$$
This shows that, if $\Ro > 1$ there exists a unique $S_\infty(p)\in(0,1)$ solution to $G(S_\infty(p),p)=0$. For future use, note that necessarily $G_x(S_\infty(p),p)< 0$.

To compute $S'_\infty(p)$, we use the implicit function theorem, obtaining
    \begin{align} \label{derivative}
        S'_\infty(p) = \dfrac{R_0 \cdot \left(S_\infty(p)-1\right)-\left(S_\infty(p)\right)^{-\tfrac{1}{p}} \cdot \log\left(S_\infty(p)\right)}
        %{p \cdot R_0 - p \cdot \left(S_\infty(p)\right)^{-1-\tfrac{1}{p}}}
        {p^2 G_x(S_\infty(p),p)}.
    \end{align}

We prove that $\eqref{derivative}$ is negative.
We need to show that the numerator is positive, since the denominator is negative. 

Considering that the first term of the numerator is negative, and the second term is positive, it is enough to show 
\begin{align*}
R_0 \cdot \left(1-S_\infty(p)\right) < - \log\left(S_\infty(p)\right) \cdot \left(S_\infty(p)\right)^{-\tfrac{1}{p}} .
\end{align*}
From $\eqref{fs}$, it follows that
\begin{align*}
    R_0 \cdot \left(S_\infty(p)-1\right) = p \cdot \left(1-\left(S_\infty(p)\right)^{-\tfrac{1}{p}}\right).
\end{align*}
The thesis is then
\begin{align*}
    p \cdot \left(\left(S_\infty(p)\right)^{-\tfrac{1}{p}}-1\right) < - \log\left(S_\infty(p)\right) \cdot \left(S_\infty(p)\right)^{-\tfrac{1}{p}},
\end{align*}
% after some manipulation and ordering:
% \begin{align*}
%   p \cdot \bigg(S_\infty(p)\bigg)^{-\tfrac{1}{p}} - p< - \log\bigg(S_\infty(p)\bigg) \cdot \bigg(S_\infty(p)\bigg)^{-\tfrac{1}{p}} .
% \end{align*}
which is equivalent to 
\begin{multline*}
  \left(p + \log\left(S_\infty(p)\right)\right) \cdot \left(S_\infty(p)\right)^{-\tfrac{1}{p}}  < p \\ \iff 
  1 + \frac{1}{p} \log\left(S_\infty(p)\right)  < \left(S_\infty(p)\right)^{\tfrac{1}{p}} \\
  \iff   1 + \log\left(S_\infty(p)\right)^{\frac{1}{p} }  < \left(S_\infty(p)\right)^{\tfrac{1}{p}}.
\end{multline*}
The final inequality follows from the well-known relation
$$ \log(y) < -(1-y), \qquad y \in (0,1),$$
with $y = \left(S_\infty(p)\right)^{\tfrac{1}{p}}$. 
\end{proof}

\medskip

\begin{proof}[Proof of Theorem \ref{theor_corr}]
Consider the function
\begin{equation}
    \label{G2p}
    F(x,p) =  \frac{p+1}{\Ro}\left(x^{-\tfrac{1}{p}} -1\right) -  (1-x^{1+\tfrac{1}{p}}).
\end{equation}
Obviously, a solution $S^*_\infty(p)$ to \eqref{eqcorr} satisfies $F(S^*_\infty(p),p)=0$. We follow the same approach as in the proof of Theorem \ref{theor_ind}.
We have
\begin{align*}
F_x(x,p) &= -\frac{p+1}{p\Ro} \left(x^{-\tfrac{1}{p}-1} -\Ro x^{\tfrac{1}{p}}\right),\\
F_{xx}(x,p) &= \frac{p+1}{p^2\Ro}\left((p+1)x^{-\tfrac{1}{p}-2}+\Ro x^{\tfrac{1}{p}-1}\right)>0.
\end{align*}
Hence $F$, as a function of $x$, is convex. Furthermore
$$
\lim_{x \to 0^+} F(x,p) = +\infty,\qquad F(1,p) = 0,\qquad F_x(1,p) =\frac{p+1}{p\Ro}(\Ro-1).
$$
This shows that, if $\Ro > 1$, there exists a unique $S^*_\infty(p)\in(0,1)$ solution to $F(S_\infty^*(p),p)=0$. For future use, note that necessarily $F_x(S^*_\infty(p),p)< 0$.

To compute ${S^*_\infty}'(p)$, we use the implicit function of theorem, obtaining
    \begin{align} \label{derivativecorr}
        {S^*_\infty}'(p) = \dfrac{p^2 \cdot {S^*_\infty}^{\tfrac{1}{p}}-p^2 -(p+1)\log {S^*_\infty} + R_0 \cdot {S^*_\infty}^{1+\tfrac{2}{p}} \cdot  \log {S^*_\infty}}
        %{R_0(p+1)p-p(p+1)\cdot {S^*_\infty}^{-1-\tfrac{2}{p}}}
        {p^2 \Ro {S^*_\infty}^{\tfrac{1}{p}}F_x(S^*_\infty(p),p)}.
    \end{align}
Since the denominator in \eqref{derivativecorr} is negative, to show that ${S^*_\infty}'(p) < 0$, we only need to prove that the numerator is positive, namely that
\begin{align} \label{prop1}
    p^2 \cdot \left({S^*_\infty}^{\tfrac{1}{p}}-1\right) > (p+1)\log {S^*_\infty} - R_0 \cdot {S^*_\infty}^{1+\tfrac{2}{p}} \cdot \log {S^*_\infty}.
\end{align}
From equation \eqref{eqcorr}, we have 
\begin{align*} 
%\label{r0corr}
    R_0 = \frac{(p+1) \left(1-{S^*_\infty}^{-\tfrac{1}{p}}\right)}{{S^*_\infty}^{1+\tfrac{1}{p}}-1}.
\end{align*}
Substituting it in \eqref{prop1} and dividing by $p^2$, one gets
% \begin{align*}
%     p^2 \cdot \left({S^*_\infty}^{\tfrac{1}{p}}-1\right) > (p+1)\log {S^*_\infty} - \frac{(p+1) \cdot \left(1-{S^*_\infty}^{-\tfrac{1}{p}}\right)}{{S^*_\infty}^{1+\tfrac{1}{p}}-1} \cdot {S^*_\infty}^{1+\tfrac{2}{p}} \cdot \log {S^*_\infty}
% \end{align*}
% holds.

% We now divide both sides with the positive $p^2$ and use the properties of the logarithm:
\begin{align*}
    \left(1-{S^*_\infty}^{\tfrac{1}{p}}\right) < - \frac{p+1}{p^2}\cdot \log \left({S^*_\infty}\right)\cdot \Bigg( 1 -  \frac{\left({S^*_\infty}^{-\tfrac{1}{p}}-1\right)\cdot {S^*_\infty}^{1+\tfrac{2}{p}}}{1-{S^*_\infty}^{1+\tfrac{1}{p}}} \Bigg).
\end{align*}
Since, as ${S^*_\infty}^{\tfrac{1}{p}} \in (0,1)$, we have 
$$ 1-{S^*_\infty}^{\tfrac{1}{p}} < -  \log \left({S^*_\infty}^{\tfrac{1}{p}}\right)=-\dfrac{1}{p}\log \left({S^*_\infty}\right),$$ it is enough showing that 
\begin{equation*}
   - \log \left({S^*_\infty}\right) > -\frac{p+1}{p}\cdot  \log \left({S^*_\infty}\right) \cdot  \Bigg( 1 -  \frac{\left({S^*_\infty}^{-\tfrac{1}{p}}-1\right)\cdot {S^*_\infty}^{1+\tfrac{2}{p}}}{1-{S^*_\infty}^{1+\tfrac{1}{p}}} \Bigg). \\
\end{equation*}
Since $\log \left({S^*_\infty}\right)$ is negative, after multiplying by $ p/(p+1)$, this is equivalent to
\begin{multline*}
    \frac{p}{p+1} <  1 -  \frac{\left({S^*_\infty}^{-\tfrac{1}{p}}-1\right)\cdot {S^*_\infty}^{1+\tfrac{2}{p}}}{1-{S^*_\infty}^{1+\tfrac{1}{p}}}\\ \iff \frac{\left({S^*_\infty}^{-\tfrac{1}{p}}-1\right)\cdot {S^*_\infty}^{1+\tfrac{2}{p}}}{1-{S^*_\infty}^{1+\tfrac{1}{p}}} <  \frac{1}{p+1} \\
    \iff (p+1) {S^*_\infty}^{1+\tfrac{2}{p}}-(p+2){S^*_\infty}^{1+\tfrac{1}{p}}+1 >0.
\end{multline*}
The final inequality is true, as shown below in Lemma \ref{lemma_ineq}; hence, we have proved \eqref{prop1}.
\end{proof}

\medskip 

\begin{lemma}
\label{lemma_ineq}
For each $p >0$ and $x \in (0,1)$, $$(p+1) x^{1+\tfrac{2}{p}}-(p+2){x}^{1+\tfrac{1}{p}}+1 >0.$$
% \begin{equation*}
%      \frac{\left(1-S_\infty^{-\tfrac{1}{p}}\right)\cdot S_\infty^{1+\tfrac{2}{p}}}{S_\infty^{1+\tfrac{1}{p}}-1} < \frac{S_\infty^{1+\tfrac{1}{p}}\cdot \log S_\infty}
%      {p \cdot \left(S_\infty^{1+\tfrac{1}{p}} -1\right)} < \frac{1}{p+1}.
% \end{equation*}
\end{lemma}
   \begin{proof}
   Let $H(x) = (p+1) x^{1+\tfrac{2}{p}}-(p+2){x}^{1+\tfrac{1}{p}}+1$.
   We have $H(1) = 0$ and
   $$ H'(x) = \frac{(p+1)(p+2)}{p}\left(x^{\tfrac{2}{p}}-x^{\tfrac{1}{p}} \right)= -\frac{(p+1)(p+2)}{p} x^{\tfrac{1}{p}}(1-x^{\tfrac{1}{p}})<0$$
   for all $x \in (0,1)$. 
   The thesis follows immediately.
   \end{proof}

   \medskip 
   
\begin{proof}[Proof of Proposition \ref{prop_compare}]
We know that $G(S_\infty(p),p) = F(S^*_\infty(p),p)=0,$ where $G$ is given in \eqref{G1p} and $F$ in \eqref{G2p}. We multiply $F$ by $\Ro/(p+1)$ to obtain 
$$
\tilde F(x,p) = x^{-\tfrac{1}{p}} -1- \frac{\Ro}{p+1} (1-x^{1+\tfrac{1}{p}}).
$$
Obviously also $\tilde F(S^*_\infty(p),p)=0$.
Therefore, we have
$$
G(1,p) = \tilde F(1,p) = 0
$$
and
$$ G_x(x,p) - \tilde F_x(x,p) = \frac{\Ro}{p}\left( 1 - x^{\tfrac{1}{p}}\right)  > 0\quad\mbox{for } x \in (0,1).$$
Hence $\tilde F(x,p) > G(x,p)$ for each $x \in (0,1)$.
The thesis follows immediately.
\end{proof}
\begin{proof}[Proof of Proposition \ref{prop_bimodal_gamma}]
Remembering $p = 1/V$, the thesis is equivalent to
\begin{equation}
    \label{thesis_gamma>bimodal}
    S_\infty(p) < \frac{1+ p \bar x}{p+1}.
\end{equation}
To prove \eqref{thesis_gamma>bimodal}, first of all notice that, manipulating \eqref{fs}, one sees that $S_\infty(p)$ can be seen as the solution in $(0,1)$ of
\begin{equation}
  \label{G_ind} G(x) := \log(x) + p \log\left(1+\frac{\Ro}{p}(1-x) \right) = 0 . 
\end{equation}
From the properties of $G(\cdot)$, it follows that \eqref{thesis_gamma>bimodal} is equivalent, using $R_0(1-\bar x) = -\log(\bar x)$, to
\begin{align}\log\left(\dfrac{1 + p\bar x}{p+1}\right) + p \log\left(1-\frac{ \log \bar x}{p+1}\right) >0.
\label{thesis2}
\end{align}
We will prove that \eqref{thesis2} holds for any $0 < \bar x <1.$
Let \begin{align*}
g(x) := \log (1 + p x) - \log (p+1) + p \log\left(1-\frac{ \log x}{p+1}\right).
\end{align*}
It is clear that $g(1) = 0.$ To prove that $g(x) >0$  for $x \in (0,1)$, we will check that $g(x)$ is decreasing in $\left(0,1\right)$. Differentiating yields  \begin{align*}
    g'(x) = p \left(\frac{1}{1+px} - \frac{1}{x (1 + p)  \left(1-\dfrac{\log x}{p+1}\right)} \right)\\ =
 p \left(\frac{1}{1+px} - \frac{1}{x + px - x \log x} \right).
\end{align*}
Considering $p >0$, we check that \begin{gather*}
    \frac{1}{1+px} < \frac{1}{x + px - x \log x}\iff
   1+px > x + px - x \log x\\ \iff
  x \log x > x - 1, 
\end{gather*} which holds in $(0,1).$
\end{proof}
\section{Supplementary figures}\label{sec:suppl_fig}
\begin{figure}[H]
    \centering
   \begin{tabular}{cc}
     \includegraphics[width=0.45\linewidth]{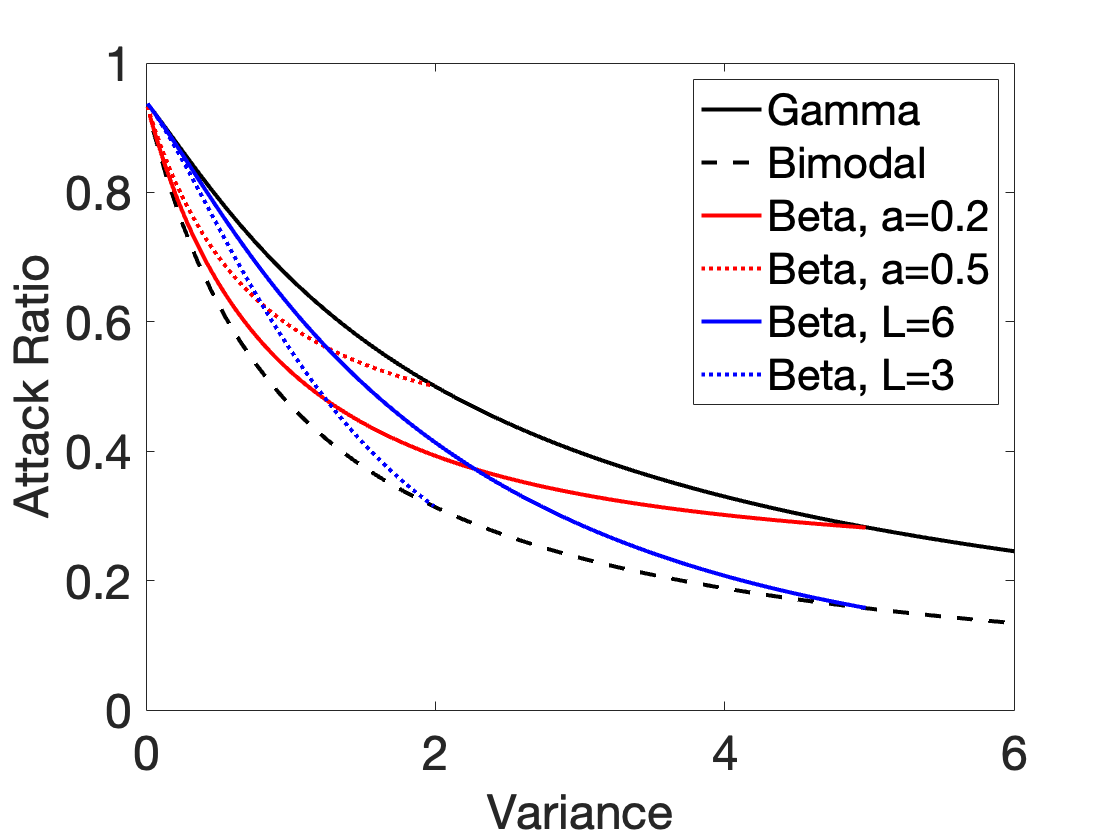}\ 
        & 
     \includegraphics[width=0.45\linewidth]{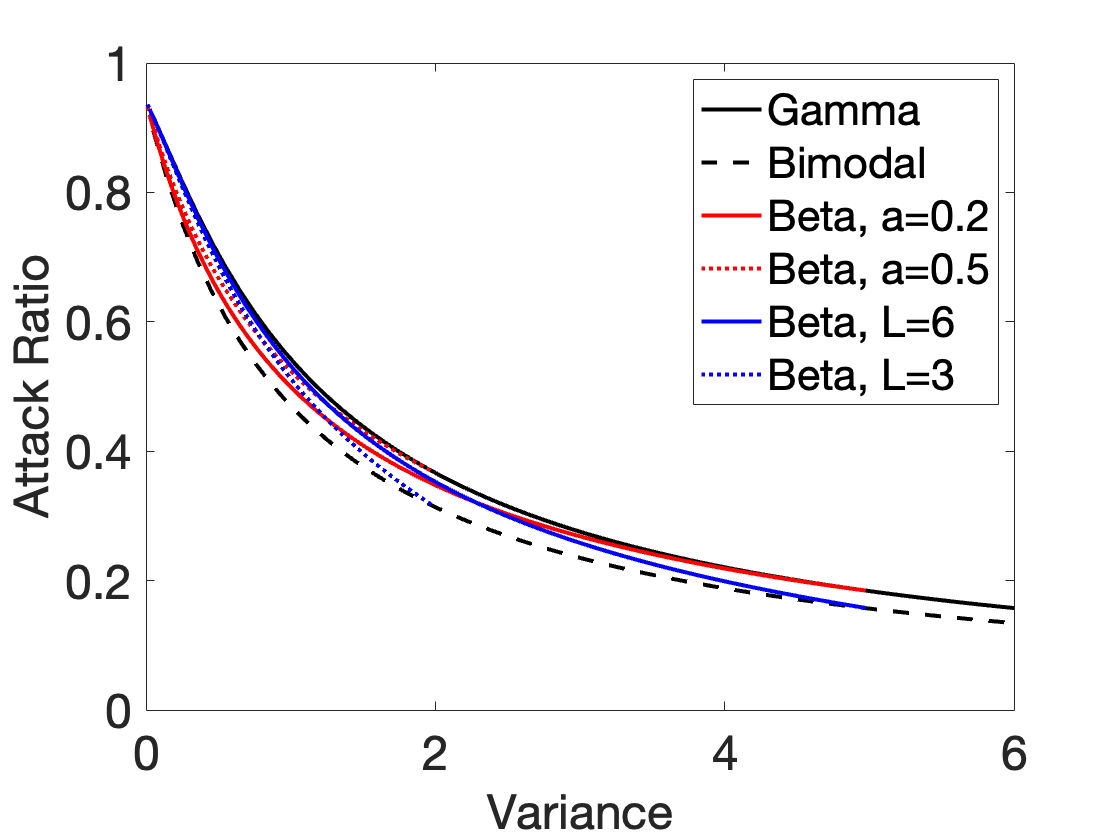}\ 
         \end{tabular}
   \caption{Left panel: Attack ratio at the end of an epidemic for system \eqref{sys_bar}, plotted as a function of the variance with $R_0=3$; curves are shown when susceptibility $X$ follows the gamma distribution, or the Beta distributions with either the parameter choices in \eqref{choice_beta1} or \eqref{choice_beta2} and when $X$ is either equal to 0 or to $V+1$ (bimodal distribution). Right panel: Same as left panel but for system \eqref{sys_bar2}--\eqref{sys_bar2b}.}
    \label{fig:finalsize_R03_ind_corr}
\end{figure}
\begin{figure}[H]
    \centering
   \begin{tabular}{cc}
     \includegraphics[width=0.45\linewidth]{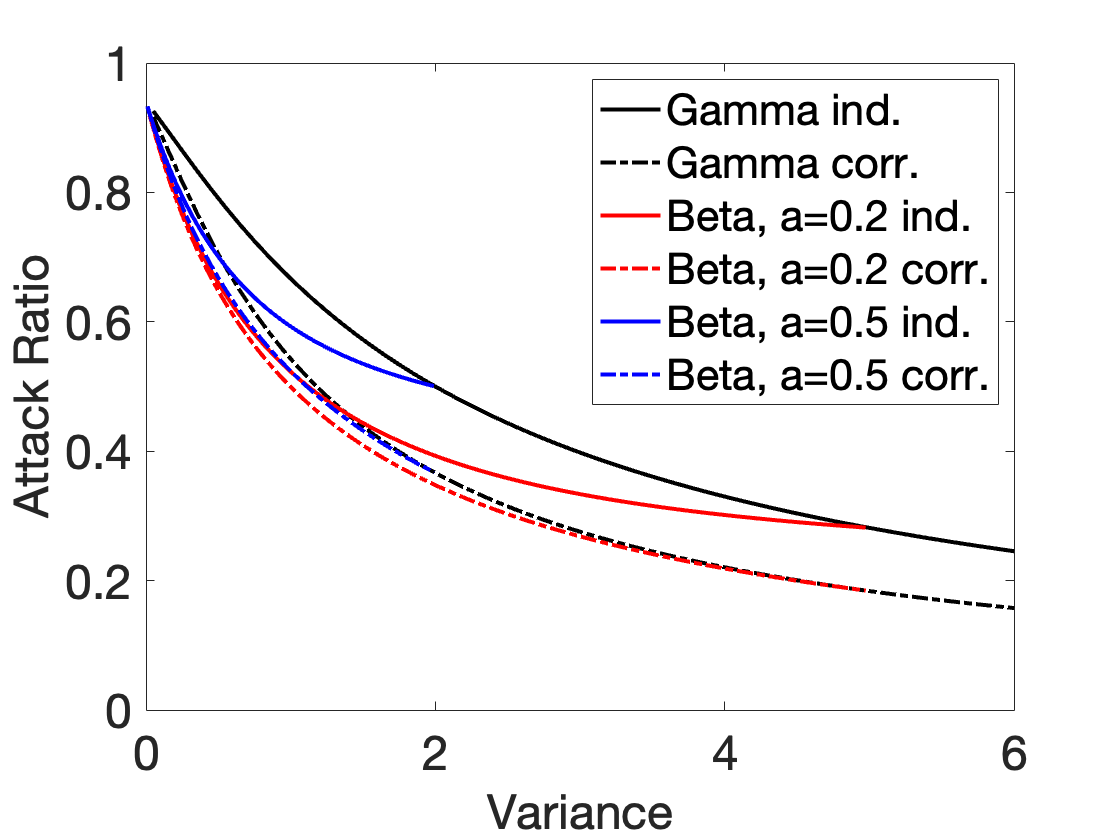}\ 
        & 
     \includegraphics[width=0.45\linewidth]{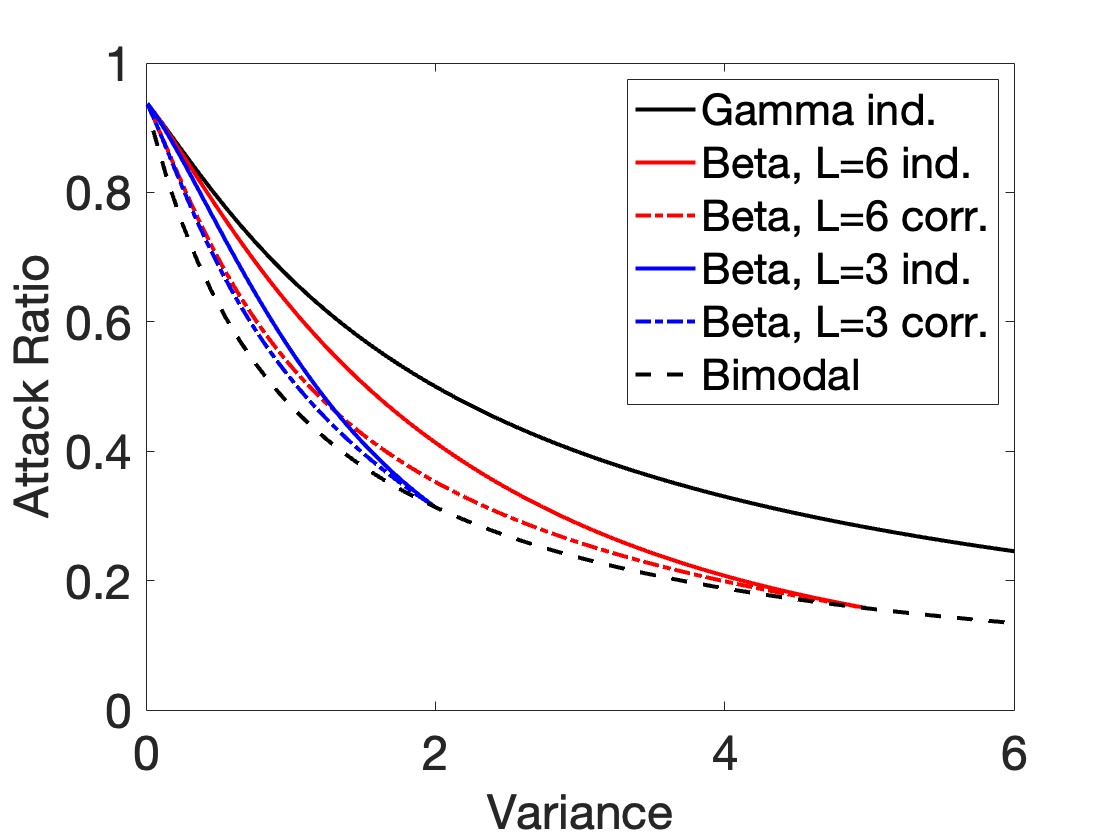}\ 
         \end{tabular}
   \caption{Left panel: Attack ratio at the end of an epidemic for systems  \eqref{sys_bar} \eqref{sys_bar2}--\eqref{sys_bar2b}, plotted as a function of the variance with $R_0=3$; curves are shown when susceptibility $X$ follows the gamma distribution, or the Beta distributions with the parameter choices in \eqref{choice_beta1}. Right panel: Attack ratio at the end of an epidemic for systems  \eqref{sys_bar} \eqref{sys_bar2}--\eqref{sys_bar2b}, plotted as a function of the variance with $R_0=3$; curves are shown when susceptibility $X$ follows the gamma distribution, or the Beta distributions with the parameter choices in \eqref{choice_beta2}.}
    \label{fig:finalsize_compare}
\end{figure}

% \begin{comment}
% \begin{thebibliography}{99}
% \bibitem{DiekmannInaba2023} O. Diekmann, H. Inaba,
% A systematic procedure for incorporating separable static heterogeneity into compartmental epidemic models,  \textit{J. Math. Biol. } \textbf{86}, 29
% (2023). \url{https://doi.org/10.1007/s00285-023-01865-0}

% \bibitem{Polymod}
% Mossong, J., Hens, N., Jit, M., Beutels, P., Auranen, K., Mikolajczyk, R., Massari, M., Salmaso, S., Tomba, G. S., Wallinga, J., Heijne, J., Sadkowska-Todys, M., Rosinska, M., Edmunds, W. J.  
% Social contacts and mixing patterns relevant to the spread of infectious diseases.  
% {\em PLoS Medicine}  
% {\bf 2008}, {\em 5(3)}.  
% \url{https://doi.org/10.1371/journal.pmed.0050074}.

% \bibitem{BrittonBallTrapman2020}
% Britton, T., Ball, F., Trapman, P.  
% A mathematical model reveals the influence of population heterogeneity on herd immunity to SARS-COV-2.  
% {\em Science}  
% {\bf 2020}, {\em 369(6505)}, 846–849.  
% \url{https://doi.org/10.1126/science.abc6810}.

% \bibitem{Novozhilov2008}
% Novozhilov2008, A. S.  
% On the spread of epidemics in a closed heterogeneous population.  
% {\em Mathematical Biosciences}  
% {\bf 2008}, {\em 215(2)}, 177–185.  
% \url{https://doi.org/10.1016/j.mbs.2008.07.010}.

% \bibitem{abramowitzstegun}
% Abramowitz, M., Stegun, I. A. (Eds.).  
% Handbook of Mathematical Functions with Formulas, Graphs, and Mathematical Tables.  
% {\em Dover Publications}  
% {\bf 1964}.

% \bibitem{covidhun}
% G. Röst, {\em et al.} 
% Early phase of the COVID-19 outbreak in Hungary and post-lockdown scenarios.
% {\em Viruses} 
% {\bf 2020}, {\em 12(7)}, 708. 
% \url{https://www.mdpi.com/1999-4915/12/7/708}.

% \end{thebibliography}
% \end{comment}
\end{document}